\documentclass[twocolumn,pre,letterpaper,amsmath,amssymb,amsfonts,floatfix,aps,
superscriptaddress,showkeys,showpacs]{revtex4-1}

\usepackage{verbatim}
\usepackage{graphicx}
\usepackage{bm} 
\usepackage{ifpdf} 
\usepackage{dcolumn}  
\usepackage{epsfig}
\usepackage{color}
\usepackage[usenames,dvipsnames]{xcolor}
\usepackage{multirow}

\newcommand{\wrho}{\widetilde{\rho}^{\,(1)}}

\newcommand{\wvbf}{\widetilde{\mathbf{v}}^{(1)}}
\newcommand{\wvl}{\widetilde{v}^{\mathrm{L}}}
\newcommand{\wvlbf}{\widetilde{\mathbf{v}}^{\mathrm{L}}}

\newcommand{\wvtbf}{\widetilde{\mathbf{v}}^{\mathrm{T}}}

\newcommand{\csch}{\operatorname{csch}}

\begin{document}

\title{Hydrodynamic fluctuation-induced forces in confined fluids}

\author{Christopher \surname{Monahan}} 
\affiliation{Department of Physics and Astronomy, University of Utah,
Salt Lake City, Utah 84112, USA}
\altaffiliation{Current address: New High Energy 
Theory Center, Rutgers, The State University of New Jersey, 136 Frelinghuysen 
Road, Piscataway, NJ 08854-8019}
\author{Ali \surname{Naji}}
\affiliation{School of Physics, Institute for Research in Fundamental Sciences 
(IPM), Tehran 
19395-5531, Iran}
\author{Ronald \surname{Horgan}}
\affiliation{DAMTP, Centre for Mathematical Sciences, University of
Cambridge, Cambridge, CB3 0WA, United Kingdom}
\author{Bing-Sui \surname{Lu}}
\affiliation{Department of Theoretical Physics, J. Stefan Institute, SI-1000 
Ljubljana, Slovenia}
\author{Rudolf \surname{Podgornik}}
\affiliation{Department of Theoretical Physics, J. Stefan Institute, SI-1000 
Ljubljana, Slovenia}
\affiliation{Department of Physics, Faculty of Mathematics and Physics, University of Ljubljana, SI-1000 Ljubljana, Slovenia}
\affiliation{Department of Physics, University of Massachusetts, Amherst, MA 01003, USA}

\begin{abstract}
We study thermal, fluctuation-induced hydrodynamic interaction forces in a 
classical, compressible, viscous fluid confined between two rigid, 
planar walls with no-slip boundary conditions. We calculate hydrodynamic 
fluctuations using the linearized, stochastic Navier-Stokes 
formalism of Landau and Lifshitz. The mean fluctuation-induced force 
acting on the fluid boundaries vanishes in this system, so we evaluate the 
two-point, time-dependent force correlations. The equal-time correlation 
function of the forces acting on a single wall gives the force variance, which 
we show to be finite and independent of the plate separation at large 
inter-plate distances. The equal-time, cross-plate force correlation, on the 
other hand, decays with the inverse inter-plate distance and is independent of 
the fluid viscosity at large distances; it turns out to be negative over the 
whole range of plate separations, indicating that the two bounding plates are 
subjected to counter-phase correlations. 
We show that the time-dependent force correlations exhibit damped temporal 
oscillations for small plate separations and a more irregular oscillatory 
behavior at large separations. The long-range hydrodynamic  correlations 
reported here represent a ``secondary Casimir effect", because the mean 
fluctuation-induced force, which represents the primary Casimir effect, is 
absent.
\end{abstract}


\maketitle

\section{Introduction}

The Casimir effect \cite{Casimir} is the most important example of a slew of phenomena 
usually referred to as {\em fluctuation-induced interactions}, their 
phenomenology extending from cosmology on the one side to 
nanoscience on the other \cite{Kardar,Dalvit,Mostepanenko,Bordag,Mohideen,Krech,
French-RMP,Parsegian2005}. The general idea tying these diverse phenomena 
together is that the confining surfaces constrain the quantum and 
thermal field fluctuations, inducing long-range interactions 
between these boundaries \cite{Kardar,Bordag}. For electromagnetic 
fields, these confinement effects lead to the Casimir-van der 
Waals interactions that can be derived within the specific framework of QED, 
and quantum field theory more generally \cite{Mohideen}. Inspired by the close 
analogy 
between thermal fluctuations in fluids and quantum fluctuations in 
electromagnetism, Fisher and de Gennes predicted the existence of long-range 
fluctuation forces in other types of critical condensed matter 
 systems \cite{Fisher} and the terms ``Casimir'' or ``Casimir-like effect'' now 
denote a range of other non-electromagnetic 
fluctuation-induced forces \cite{Kardar,French-RMP}. 

Beyond detailed measurements of the Casimir-van der Waals interactions 
\cite{Mohideen}, attention has been directed 
toward Casimir-like forces engendered by density fluctuations in the 
vicinity of the vapor-liquid critical point \cite{Krech,Krech2,Krech3};
in binary liquid mixtures near the critical demixing point 
\cite{Gambassi,Fukuto}; and in thin polymer \cite{Morariu04,Schaeffer,Morariu03} 
and liquid crystalline films \cite{LC,Li-kardar}.
Most recently, several studies have examined fluctuation-induced interactions 
for the Casimir-Lifshitz force out of thermal equilibrium \cite{Antezza,Kruger}, 
for the temporal relaxation of the thermal Casimir or van der Waals force 
\cite{Dean}, and for nonequilibrium steady states in 
fluids \cite{Kirkpatrick13,Kirkpatrick14,Ortiz}, where fluctuations 
are anomalously large and long-range.

It is instructive to 
recall that the original 1955 derivation of the electromagnetic Casimir-van der 
Waals interactions by Lifshitz \cite{Lifshitz} was not 
fundamentally rooted in QED but rather in stochastic electrodynamics, first 
formulated by Rytov \cite{rytov59}. In stochastic 
electrodynamics, Maxwell's equations are augmented by fluctuating displacement 
current sources \cite{Rosa}. This leads to two coupled electrodynamic 
Langevin-type equations, for each of the fundamental electrodynamic fields, that are 
then solved with standard boundary conditions. The interaction force is 
obtained by averaging the Maxwell stress tensor and taking into account the 
statistical properties of the fluctuating sources \cite{Narayanaswami}. This 
paradigmatic Lifshitz-route to fluctuation-induced interactions later became 
disfavored as other formal approaches gained strength \cite{Mohideen}, 
but appears to be reborn in recent endeavors regarding non-equilibrium 
fluctuation-induced interactions \cite{Kirkpatrick13,Kirkpatrick14}. In fact, in the 
Dean-Gopinathan method there exists a mapping of the non-equilibrium problem 
characterized by dissipative dynamics onto a corresponding static (Lifshitz) 
partition function provided by the Laplace transform of the time-dependent force 
and the static partition function \cite{Ajay1,Ajay2}.

Based on the success of stochastic electrodynamics, Landau and Lifshitz 
proposed by analogy the stochastic dissipative hydrodynamic equations 
\cite{LandauLifshitz}, augmenting the linearized Navier-Stokes equations with 
fluctuating heat flow vector and fluctuating stress tensor \cite{Ortiz,Forster}. 
This leads to three coupled hydrodynamic equations involving the fundamental 
hydrodynamic fields of mass density, velocity and local temperature, which can 
now be solved in different contexts. In the absence of 
thermal conductivity, this system further reduces to a Langevin-type equation 
for the velocity 
field, involving the stress tensor fluctuations, and a continuity equation for 
the mass density field. Since the fundamental hydrodynamic equations are 
non-linear, the derivation of fluctuating Landau-Lifshitz hydrodynamics already 
involves heavy linearity {\em Ans\" atze} and the possible generalization to a 
full non-linear fluctuating hydrodynamics is not clear \cite{spohn,Tailleur}.

Although fluctuating electrodynamics is based on 
{\em linear} Maxwell's equations with stresses {\em quadratic} in the field and
fluctuating hydrodynamics stems from {\em non-linear} Navier-Stokes equations 
with stresses {\em linear} in velocities, the general similarity between these 
approaches might nevertheless lead one to assume that, in confined geometries, there 
should exist Casimir-like hydrodynamic fluctuation forces. But this notion is 
at odds with the standard decomposition of the classical partition function into 
momentum and configurational 
parts. This decomposition has far-reaching consequences, 
which were clearly understood as far back as van der Waals' 
thesis \cite{Rowlinson}. While there is an analogy between the description of 
fluctuations in these two areas of physics, caution should be exercised when 
trying to translate results from one field directly into the other. We will show 
that there does exist a type of Casimir effect in the hydrodynamic 
context, but that this effect has fundamentally different properties from the 
conventional Casimir effect.

The first step in bringing together the Casimir force in electrodynamics and 
its putative counterpart in hydrodynamics was made by Jones \cite{Jones}. 
Inspired by the obvious analogy between electrodynamics and hydrodynamics, Jones
investigated the possible existence of a long-ranged, fluctuation-induced, 
effective force generated by confining boundaries in a fluid. He showed that  
in linearized hydrodynamics the net (mean) stochastic force vanishes, which 
led him to introduce a next-to-leading order formalism. 
The status of this formalism, however, is not entirely clear, 
because there are linearity assumptions rooted deep within fluctuational 
hydrodynamics \cite{Forster,Ortiz}. Within the context of this next-to-leading 
order formalism, Jones 
demonstrated that long-range forces could exist in a semi-infinite fluid
or around an immersed spherical body, and would be strongest in 
incompressible fluids, with much weaker forces in compressible fluids. This 
result is at odds with the momentum decomposition of the classical partition 
function and should be considered an artifact of the next-to-leading order 
analysis of the stochastic equations governing the hydrodynamic field 
evolution. 

Chan and  White \cite{Chan}, therefore, reconsidered the whole calculation. They 
concentrated on the planar geometry of two hard walls immersed in a fluid and 
argued that hydrodynamic fluctuations could give rise to a repulsive force in 
{\em incompressible} fluids, but that this force would vanish for classical 
{\em compressible} fluids. Since an incompressibility {\em Ansatz} 
does not translate directly into the interaction potential in the classical
partition function \cite{vankampen}, this fictional case could lead to a 
fluctuation-induced interaction that would not be contrary to the argument based on the momentum 
decomposition of the classical partition function. The repulsive fluctuation-induced 
force would also in itself not be that hard to envision since the existence of 
a repulsive force in the context of van der Waals  interactions is 
well-established and was originally proposed in Ref.~\cite{Dzyaloshinskii}. The 
vanishing of the fluctuation-induced force for classical compressible fluids is based on 
a rough argument of analytic continuation of the viscosities into the infinite 
frequency domain \cite{Chan}. While this latter argument is appropriate in 
electrodynamics, because an infinite frequency corresponds to the vacuum, it is 
not reasonable in hydrodynamics, where the whole basis of the continuum 
hydrodynamic theory breaks down before any such limit could be enforced 
\cite{Forster}.

Therefore, both approaches to the problem of hydrodynamic Casimir-like interactions 
have strong limitations and subsequent developments  
failed to conclusively prove either point of view \cite{Ivlev}.

In this paper, we revisit the question of the existence of long-range, 
fluctuation-induced forces in classical fluids. We work strictly within the 
framework of 
linearized stochastic hydrodynamics and rather than considering the net force, 
which is zero trivially, we study the force correlators.
In other words, we focus on the question: In what way do boundary conditions 
and statistical properties of the fluctuating hydrodynamic stresses affect the 
statistical properties (correlators) of the random forces acting on the bounding surfaces?

We formulate a general approach to this problem by considering a fluid of 
arbitrary compressibility, bounded 
between two plane-parallel, hard walls with 
no-slip boundary conditions. Thermal fluctuations lead to spatio-temporal 
variations in the pressure and velocity fields that can be calculated 
using the linearized, stochastic Navier-Stokes formalism of Landau and Lifshitz
\cite{LandauLifshitz}. Within this approach, we derive analytical expressions 
for the time-dependent correlators  (for both the same-plate and the
cross-plate correlators)
of the fluctuation-induced forces acting on the walls. 
In particular, we express the variance of these forces in terms of frequency 
integrals that have simple plate-separation dependence in the 
small and large plate-separation limits.

Our results do not depend upon the 
next-to-leading order formalism of Jones \cite{Jones}, nor do they depend on the
unrealistic validity of analytic continuation of the viscosities in the whole 
frequency domain \cite{Chan}. We show that, while the mean force vanishes, 
the variance of the fluctuation-induced normal force is finite and 
depends on the separation between the bounding surfaces. We call this the 
\emph{secondary Casimir 
effect}, because the primary Casimir effect refers to the average value of 
the fluctuation-induced force (which is zero here) and not strictly its variance. Both quantities 
have been investigated in other Casimir-like situations \cite{Bartolo, Dean} 
and in disordered charged systems \cite{disorder-PRL,pre2011,epje2012}.
The equal-time, 
cross-plate force correlation exhibits long-range behavior that is 
independent of the fluid viscosity and decays proportional to the inverse plate
separation. Finally, we find that the time-dependent correlators exhibit 
damped oscillatory behavior for small plate separations that becomes
irregular at large distances.

In Sec.~\ref{sec:formalism}, we outline the stochastic formalism of Landau 
and Lifshitz and the strategy of our calculation of hydrodynamic 
fluctuation-induced forces in the general case of compressible fluids. Sections 
\ref{sec:meanforce} and \ref{sec:forcevar_t} present the main steps of our 
calculation. We show results for the equal-time force correlators and the 
two-point, 
time-dependent correlators in Sections \ref{sec:numerics} and 
\ref{sec:tnumerics}, respectively. We conclude our discussions in Sec.~
\ref{sec:conclusions}.

\section{Formalism}
\label{sec:formalism}

We consider the hydrodynamic fluctuations in a Newtonian fluid at rest and in
the absence of heat transfer. These fluctuations are described by the 
stochastic 
Landau-Lifshitz equations \cite{LandauLifshitz}
\begin{align}
&\eta \nabla^2 \mathbf{v} + \left( \frac{\eta}{3} + \zeta\right) 
\nabla(\nabla\cdot \mathbf{v}) - \nabla p {}  \nonumber\\
&\qquad\qquad\qquad{}- \rho\left(\frac{\partial
\mathbf{v}}{\partial t} + \mathbf{v}\cdot \nabla\mathbf{v}\right)  
=
-\nabla\cdot \mathbf{S},  \label{eq:ns1}\\
&\frac{\partial \rho}{\partial t} + \nabla \cdot (\rho \mathbf{v}){} = 
0, \label{eq:ns2}
\end{align}
where $\mathbf{v}=\mathbf{v}(\mathbf{r};t)$, $p=p(\mathbf{r};t)$ and
$\rho=\rho(\mathbf{r};t)$ are the  velocity, pressure and
density fields and $\eta$ and $\zeta$ are the shear and bulk
viscosity coefficients, respectively \cite{Note1}. The
randomly fluctuating microscopic degrees of freedom are driven by the
random stress tensor $\mathbf{S}=\mathbf{S}(\mathbf{r};t)$, which is assumed to 
have a Gaussian distribution with zero mean $\left\langle S_{ij}(\mathbf{r}; t) 
\right\rangle = {}  0$ and the two-point correlator 
\begin{align}
&\left\langle
S_{kl}(\mathbf{r};t)\,S_{mn}(\mathbf{r}
^{\prime};t^{\prime}) \right\rangle= {} 
2k_{\mathrm{B}}T\delta(\mathbf{r}-\mathbf{r}^{\prime})\delta(t -
t^{\prime}) \nonumber\\
&\qquad \times \Big[
 \eta\big(\delta_{km}\delta_{ln} +
\delta_{kn}{} \delta_{lm}\big)
- \left( \frac{2\eta}{3}-\zeta \right) \delta_{kl}
\delta_{mn} \Big].  \label{eq:gaussprop2} 
\end{align}
Here the subindices ($i, j, k, \ldots$) denote the Cartesian components $(x, 
y, z)$,  $k_{\mathrm{B}}$ is Boltzmann's constant and $\langle\cdots \rangle$ 
denote an equilibrium ensemble average at temperature $T$. We do not consider 
any possible relaxation effects, which would formally correspond to 
frequency-dependent viscosities, but these effects can be easily incorporated 
\cite{LandauLifshitz}. Denoting the frequency Fourier 
transform by a tilde, i.e., 
\begin{equation}
\widetilde{f}(\omega) = \int \mathrm{d}t\,e^{i\omega t} f(t),
\end{equation}
we have $\langle \widetilde{S}_{ij}(\mathbf{r};\omega) \rangle 
=   0$ and 
\begin{align}
&\left\langle
\widetilde{S}_{kl}(\mathbf{r};\omega)\,\widetilde{S}_{mn}^{\ast}(\mathbf{r}
^{\prime};\omega^{\prime}) \right\rangle= {} 
4\pi k_{\mathrm{B}}T\delta(\mathbf{r}-\mathbf{r}^{\prime})\delta(\omega -
\omega^{\prime}) \nonumber\\
&\qquad \times \Big[
 \eta\big(\delta_{km}\delta_{ln} +
\delta_{kn}\delta_{lm}{} \big)
- \left( \frac{2\eta}{3}-\zeta \right) \delta_{kl}
\delta_{mn} \Big],  \label{eq:gausswprop2} 
\end{align}
which hold independent of the boundary conditions imposed on the fluid system.

Before proceeding further, we should note that this form of fluctuating 
hydrodynamics is analogous to the
Rytov fluctuating electrodynamics \cite{Lifshitz}, where the basic equations 
for the electric and magnetic fields are 
\begin{align}
\nabla \times \mathbf{E}(\mathbf{r}, t) = {} & -  \frac{\partial 
\mathbf{B}(\mathbf{
r}, t)}{\partial t}, \label{bcfhgjsk}\\
\nabla \times \mathbf{H}(\mathbf{r}, t) = {} & \frac{\partial 
\mathbf{D}(\mathbf{ r}, 
t)}{\partial t} + \frac{\partial \mathbf{K}(\mathbf{r}, t)}{\partial t}, 
\end{align}
supplemented by $\nabla \cdot \mathbf{D}(\mathbf{r}, t) = 0$ and  $\nabla \cdot 
\mathbf{B}(\mathbf{r}, t) = 0$ and appropriate boundary conditions. In this 
case, the fluctuating random polarization, $\mathbf{K}(\mathbf{r}; t)$, 
has Gaussian properties with $\big\langle \widetilde{K}_{i}(\mathbf{r};\omega) 
\big\rangle =  0$, and 
\begin{equation}
\left\langle  
\widetilde{K}_{i}(\mathbf{r};\omega)\,\widetilde{K}_{j}^{\ast}(\mathbf{r}
^{\prime};\omega^{\prime}) \right\rangle = {} 
{k_{\mathrm{B}}T}\frac{\varepsilon_{\mathrm{I}}(\omega)}{\omega}\delta_{ij} 
\delta(\mathbf{r}-\mathbf{r}^{\prime})\delta(\omega -
\omega^{\prime}),
\label{axrmeoi}
\end{equation}
where we have assumed a dispersive dielectric response function 
$\varepsilon(\omega) = 
\varepsilon_{\mathrm{R}}(\omega) + 
i \varepsilon_{\mathrm{I}}(\omega)$. We can immediately see the similarity 
between Eqs.~\eqref{eq:ns1}-\eqref{eq:gausswprop2} and 
Eqs.~\eqref{bcfhgjsk}-\eqref{axrmeoi}.

Thus, the stochastic approach to hydrodynamics is very close to 
Lifshitz's 
original analysis of the electromagnetic 
problem \cite{Lifshitz}, provided one fully 
takes into account the basic differences between the Maxwell equations and the 
Navier-Stokes equations \cite{Chan}: The former are linear in the fields with 
stresses quadratic in the fields, while the latter are non-linear in the fields 
with stresses linear in the fields. This difference leads to 
some important distinctions and precludes directly applying results from 
electrodynamics to the hydrodynamic domain.

\subsection{Linearized stochastic hydrodynamics}
\label{sec:linhydro}

For vanishing random stress tensor, the equilibrium solution of 
Eqs.~\eqref{eq:ns1} and \eqref{eq:ns2} is $\mathbf{v}=\mathbf{0}$, $p=p_0$ and 
$\rho 
=
\rho_0$, corresponding to a fluid at rest at constant temperature, $T$, with
uniform pressure, $p_0$, and density, $\rho_0$. The random stress tensor,
$\mathbf{S}$, is of order $k_{\mathrm{B}}T$ and, consequently, 
macroscopically small. Thus, the corresponding fluctuations in the 
velocity, pressure and density fields are also macroscopically small. 
Therefore we introduce a linearized treatment of the Landau-Lifshitz
equations, by setting $\mathbf{v} = \mathbf{v}^{(1)}$, $p
= p_0+p^{(1)}$ and $\rho = \rho_0+\rho^{(1)}$, where the superscript
$(1)$ denotes a term of order $\mathbf{S}$.

We assume local equilibrium, which enables us to relate the
density and pressure as 
\begin{equation}\label{eq:prho}
p^{(1)} = c_0^2\rho^{(1)},\qquad \mathrm{with}\qquad 
c_0^2 = \left(\frac{\partial p }{\partial \rho}\right)_0. 
\end{equation}
Here $c_0$ is the adiabatic speed of sound, so that $\rho_0 c_0^2$ equals 
the inverse adiabatic compressibility (Newton-Laplace equation). 
Eqs.~\eqref{eq:ns1} and 
\eqref{eq:ns2} can be linearized as  
\begin{align}
&\eta \nabla^2 \mathbf{v}^{(1)} + \left( \frac{\eta}{3} + \zeta\right)
\nabla\left(\nabla\cdot \mathbf{v}^{(1)}\right)    \nonumber\\
&\qquad\qquad\qquad {}- \nabla 
p^{(1)}- \rho_0\frac{\partial
\mathbf{v}^{(1)}}{\partial t}   = 
-\nabla\cdot \mathbf{S},  \label{eq:lns1}\\
&\frac{\partial \rho^{(1)}}{\partial t} + \rho_0\nabla \cdot \mathbf{v}^{(1)} 
{} 
= 0. \label{eq:lns2}
\end{align}
or, in the frequency domain and using Eq.~\eqref{eq:prho}   \cite{Note2},
\begin{align}
&\eta \nabla^2\wvbf +
\left(\frac{\eta}{3} + \zeta\right)\nabla \left(\nabla \cdot
\wvbf\right) 
{}  \nonumber\\
 &\qquad\qquad\qquad {} {} -c_0^2\nabla \wrho+ i\omega \rho_0 
\wvbf = 
-\nabla \cdot \widetilde{\mathbf{S}}, \label{eq:linLL1}\\
&\nabla \cdot \wvbf -\frac{i\omega}{\rho_0}
\wrho {} =  0.\label{eq:linLL2}
\end{align}

We now introduce transverse and longitudinal components of the velocity 
fluctuations $\mathbf{v}^{(1)}$, which 
we denote $\mathbf{v}^\mathrm{T}$ and $\mathbf{v}^\mathrm{L}$, respectively. We 
have dropped the superscript $(1)$ for notational simplicity, i.e., 
$\mathbf{v}^{(1)} = \mathbf{v}^\mathrm{T}+\mathbf{v}^\mathrm{L}$, with 
\begin{equation}
\nabla\cdot \mathbf{v}^\mathrm{T} = 0 \qquad \mathrm{and}\qquad \nabla\times 
\mathbf{v}^\mathrm{L} = 0.
\label{eq:def_LT}
\end{equation}
The random force density vector $\mathbf{\Sigma} = -\nabla \cdot \mathbf{S}$ can 
be 
decomposed into transverse and 
longitudinal components as well, using $\mathbf{\Sigma} = 
\mathbf{\Sigma}^\mathrm{T} + \mathbf{\Sigma}^\mathrm{L}$, where
\begin{equation}
\nabla\cdot \mathbf{\Sigma}^\mathrm{T} = 0 \qquad \mathrm{and}\qquad 
\nabla\times 
\mathbf{\Sigma}^\mathrm{L} = 0.
\end{equation}
These random force density vector components have zero mean and zero cross 
correlations. Their self-correlations follow 
from Eq.~\eqref{eq:gaussprop2} as
\begin{align}
\left\langle
\widetilde{\Sigma}^{\mathrm{L}}_i(\mathbf{r};\omega)\,\widetilde{\Sigma}^{
\mathrm{L}}_j(\mathbf{r}^{\prime};\omega^{\prime}) \right\rangle= {} & 
4 \pi k_{\mathrm{B}}T
\left( \frac{4\eta}{3}+\zeta \right) 
\nabla_i\nabla_j^\prime 
\nonumber\\
 & \quad \times \delta(\mathbf{r}-\mathbf{r}^{\prime})\delta(\omega +
\omega^{\prime}), \label{eq:svecLw-2}\\
\left\langle
\widetilde{\Sigma}^{\mathrm{T}}_i(\mathbf{r};\omega)\,\widetilde{\Sigma}^{
\mathrm{T}}_j(\mathbf{r}^{\prime};\omega^{\prime}) \right\rangle= {} & 
4 \pi k_{\mathrm{B}}T
\eta\left(
\nabla_k\nabla_k^\prime \delta_{ij} -
\nabla_i\nabla_j^\prime
\right)\nonumber\\
 & \quad \times \delta(\mathbf{r}-\mathbf{r}^{\prime})
\delta(\omega +
\omega^{\prime}). \label{eq:svecTw-2}
\end{align}
The stochastic Landau-Lifshitz equations can thus be written as
\begin{align}
&\eta \nabla^2\wvlbf  + \left(\frac{\eta}{3} + 
\zeta\right)\nabla\left(\nabla \cdot
\wvlbf\right) {}  \nonumber\\
&\qquad\qquad\qquad-c_0^2\nabla \wrho+ i\omega \rho_0 \wvlbf {}  = 
 \mathbf{\widetilde{\Sigma}}^\mathrm{L}, \label{eq:lllong}\\
&\eta \nabla^2\wvtbf+ i\omega \rho_0 
\wvtbf{}  = 
\mathbf{\widetilde{\Sigma}}^\mathrm{T}, \label{eq:lltrans}\\
&\nabla \cdot \wvlbf -\frac{i\omega}{\rho_0}\wrho{}  =  
0.\label{eq:lllong2}
\end{align}

We may simplify Eqs.~\eqref{eq:lllong}-\eqref{eq:lllong2} by 
using the vector identity
\begin{equation}\label{eq:vecid}
\nabla_j\nabla_j \wvl_i =  \nabla_j\nabla_j \wvl_i + 
\nabla_j\left(\nabla_i \wvl_j -\nabla_j
\wvl_i\right)= \nabla_i \nabla_j\wvl_j
\end{equation}
for the curl-free longitudinal component and by substituting 
Eq.~\eqref{eq:lllong2} into Eq.~\eqref{eq:lllong} to obtain
\begin{align}
&\left[\frac{4\eta}{3} + 
\zeta 
+ \frac{i\rho_0c_0^2}{\omega}\right]\nabla^2
\wvlbf+ i\omega \rho_0 
\wvlbf = 
{} 
 \widetilde{\mathbf{\Sigma}}^\mathrm{L}, \label{eq:ll}\\
&\eta \nabla^2\wvtbf+ i\omega \rho_0 
\wvtbf = {} 
\widetilde{\mathbf{\Sigma}}^\mathrm{T}. \label{eq:tt}
\end{align}

We have now decoupled the transverse and longitudinal components of the 
velocity fluctuations. Eqs.~\eqref{eq:ll} and \eqref{eq:tt} are
nothing but the Langevin equations for each component of the velocity field in 
the frequency domain. In fact, Eq.~\eqref{eq:ll} is a scalar equation for 
the longitudinal component of the velocity fluctuation \cite{Forster}. 

The density field fluctuations can be obtained from the longitudinal component 
of the velocity field fluctuations, 
\begin{equation}
\wrho(\mathbf{r};\omega) = -\frac{i\rho_0}{\omega}\nabla \cdot 
\wvlbf(\mathbf{r};\omega). \label{eq:rhovl}
\end{equation}

\section{Mean interaction force}
\label{sec:meanforce}

To obtain the net effective interaction force between the fluid's 
confining boundaries, we integrate the fluctuating hydrodynamic stress tensor, 
$\sigma_{ij} 
= 
\sigma_{ij}(\mathbf{r};t)$, over the bounding surfaces, 
$\Gamma$, i.e., 
\begin{equation}
\big\langle{\cal F}_i(t)\big\rangle = \int_{\Gamma}
\big\langle\sigma_{ij}(\mathbf{r};t)\big\rangle\, \mathrm{d} A_j, 
\end{equation}
where the fluctuating hydrodynamic stress tensor, which is \cite{LandauLifshitz}
\begin{equation}
\sigma_{ij} = \eta \big[\nabla_i v_j + \nabla_j v_i\big] -
\left[\left(\frac{2\eta}{3} - \zeta\right) \nabla_k v_k +
p\right]\delta_{ij}+S_{ij},
\end{equation}
 can be written up to first order in the field fluctuations as 
 $\sigma_{ij} = -p_0\delta_{ij}+\sigma_{ij}^{(1)}$, with 
\begin{align}\label{eq:sigma}
\sigma_{ij}^{(1)} = {} & \eta \big[\nabla_i v^{(1)}_j + \nabla_j v^{(1)}_i\big]
\nonumber\\
{} & \quad 
-
\left[\left(\frac{2\eta}{3} - \zeta\right) \nabla_k v_k^{(1)} 
+c_0^2\rho^{(1)}\right]\delta_{ij}+S_{ij}. 
\end{align}

The stress tensor is linear in the fluid fluctuations, which are 
themselves linear in the random stress tensor and, thus, their ensemble 
averages vanish  $\big\langle \mathbf{v}^{\mathrm{T}}(\mathbf{r};t) \big\rangle 
= \big\langle 
\mathbf{v}^{\mathrm{L}}(\mathbf{r};t) \big\rangle = 0$ and $\big\langle 
\rho^{(1)}(\mathbf{r};t)  \big\rangle =  0$. 
As a result, at first order in field fluctuations, the net fluctuation-induced force acting on the 
fluid boundaries must vanish, 
irrespective of the geometry of the fluid system, i.e., 
\begin{equation}
\big\langle{\cal F}_i^{(1)}(t)\big\rangle = 0. 
\end{equation}
We note that the mean force at leading order stems from the equilibrium 
pressure and is simply ${\cal F}_z^{(0)} = -p_0 A$. We exclude this 
contribution in the rest of our discussion and focus on the statistical 
properties of the force at first order in the field fluctuations.

In what follows, we limit our discussion to the plane-parallel geometry of two 
rigid walls of arbitrarily large surface area, $A$. We assume that the walls 
are located along the $z$ axis at $z=0$ and $z=L$ at a separation distance of 
$L$ and that the fluid velocity satisfies no-slip boundary conditions on the 
walls. 

\section{\label{sec:forcev}Two-point, time-dependent correlations of the force}
\label{sec:forcevar_t}

Although, as we have already noted, the mean inter-plate force due to 
hydrodynamic fluctuations in the fluid layer must vanish, its variance or 
correlation functions need not and do not. In 
this Section, we study the two-point, time-dependent correlators, including the 
variance, 
of the forces that act on the boundaries in the two-wall geometry.  
In this plane-parallel geometry, we are primarily  
concerned with the force perpendicular to the plane boundaries, in which case 
the two-point, time-dependent force correlator is given 
by
\begin{align}\label{eq:forcecorr}
{\cal C} {} &(z,z';t,t') =  \big\langle {\cal F}_z^{(1)}(z;t) 
{\cal F}_z^{(1)}(z';t') 
\big\rangle \nonumber\\
{} & \quad= \iint_A
\left\langle
\sigma_{zz}(\mathbf{r};t)\sigma_{zz}(\mathbf{r}^{\prime};t')\right\rangle\,
\mathrm{d} x \, \mathrm{d} y\, \mathrm{d} x'\, \mathrm{d}y', 
\end{align}
where the integrals run over the surface areas $A$ of the two walls that are 
located
at $z=0$ and $z=L$.  Throughout this paper, we use an uppercase ${\cal C}$ to 
denote  correlation 
functions of the normal forces acting on the fluid boundaries and a lowercase 
$c$ to refer 
to correlation functions 
of the fluctuating hydrodynamic fields. We express the former quantity in 
terms of the latter ones (see Appendix \ref{app:f3f3simple}). In the present 
case, the correlators of 
the velocity and density fluctuations are given 
by 
\begin{align} \label{eq:vvstart}
c_{ij}^{\mathrm{T}\mathrm{T}}(\mathbf{r},\mathbf{r}';t,t') = {} & \big\langle 
v_i^{\mathrm{T}}(\mathbf{r};t) 
v_j^{\mathrm{T}}(\mathbf{r}^\prime;t') \big\rangle,\\ 
c_{ij}^{\mathrm{L}\mathrm{L}}(\mathbf{r},\mathbf{r}';t,t') = {} & \big\langle 
v_i^{\mathrm{L}}(\mathbf{r};t) 
v_j^{\mathrm{L}}(\mathbf{r}^\prime;t') \big\rangle, \\
c^{\rho\rho}(\mathbf{r},\mathbf{r}';t,t') = {} & \big\langle 
\rho^{(1)}(\mathbf{r};t)  
\rho^{(1)}(\mathbf{r}^\prime;t')  
\big\rangle.
\end{align}
The cross-correlation function of the transverse and longitudinal 
components of the velocity
vanishes by construction. Furthermore, the transverse velocity and density 
fluctuations  are  independent fields, with vanishing cross-correlation 
function. 
Therefore, the only 
other correlation function we need is the density-velocity cross-correlator,
\begin{equation}\label{eq:vpstart}
c_i^{\mathrm{L}\rho}(\mathbf{r},\mathbf{r}';t,t')= \big\langle 
v_i^{\mathrm{L}}(\mathbf{r};t) 
\rho^{(1)}(\mathbf{r}^\prime;t')\big\rangle.  
\end{equation}

Not all of these correlators contribute to the time-dependent correlator of the 
forces between the two hard boundaries. 
In Appendices \ref{app:vtzero} and  \ref{app:pvzero}, we show that the 
contributions to the normal 
force correlator generated by the correlation function of 
the transverse velocity field and by the correlation function between the 
velocity and density fields vanish for our geometry.  
Therefore, applying the formulae of the previous Section, we can write the 
time-dependent force correlator as the sum of three terms (see  Appendix 
\ref{app:f3f3simple} for details),
\begin{equation}\label{eq:czzpt}
{\cal C}(z,z';t,t')  =  \sum_{i = 0}^2{\cal P}_i(z,z';t,t').
\end{equation}
Defining the dimensionless parameter
\begin{equation}
\chi = 4/3+\zeta/\eta,\label{eq:chi}
\end{equation}
we can write the first term as
\begin{equation}\label{eq:p0delta}
{\cal P}_0(z,z';t,t') \equiv 2k_{\mathrm{B}}T \eta \chi A 
\delta(z-z')\delta(t-t').
\end{equation}
This contribution stems directly from the integration of the random 
stress correlator, $\langle
S_{zz}(\mathbf{r};t)S_{zz}(\mathbf{r}^{\prime};t')\rangle$, over the bounding 
surfaces; this term vanishes unless $z= z'$ and $t= t'$, in which case it 
reduces to an irrelevant constant that will be dropped in the rest of our
analysis. The two other terms are
\begin{align}
{\cal P}_1(z, z'; t,t') \equiv {} &\left(\frac{4\eta}{3} + 
\zeta\right)^2\iint_A\mathrm{d} x \, \mathrm{d} y\, \mathrm{d} x'\, 
\mathrm{d} y' \nonumber\\
{} & \qquad \times 
\nabla_z\nabla_z'c^{\mathrm{L}\mathrm{L}}_{zz}(\mathbf{r},\mathbf{r}';t,
t'),\label{eq:p1t}\\
{\cal P}_2(z, z'; t,t') \equiv {} &c_0^4\iint_A\mathrm{d} x \, \mathrm{d} y\, 
\mathrm{d} x'\, \mathrm{d} y' c^{\rho\rho}(\mathbf{r},\mathbf{r}';t,
t'). \label{eq:p2t}
\end{align}
We note that, in the above, we have used Eq.~\eqref{eq:rhovl}, which relates 
the density fluctuations to the fluctuations of the longitudinal components of 
the velocity.

With this expression in hand, we can see that we need to determine the 
correlation functions of the density fields and the longitudinal component of 
the velocity fields. We proceed via the following steps \cite{LandauLifshitz}:
\begin{enumerate}
\item Obtain the Green functions of Eq.~\eqref{eq:ll};
\item Express the  fluctuating fields and their correlation functions in
terms of the Green functions above;
\item Integrate the resulting expressions over the boundaries of the
fluid according to Eqs.~\eqref{eq:p1t} and \eqref{eq:p2t}. 
\end{enumerate}

\subsection{Green functions}

In the present model with no-slip walls, the velocity and, therefore, the 
corresponding Green function
should vanish at the boundaries. Translational invariance in the two 
(transverse) directions
perpendicular to the $z$-axis prompts us 
to search for Green functions of
the form
\begin{equation}\label{eq:Gdef}
\widetilde{G}(\mathbf{r},\mathbf{r}^{\prime\prime};\omega) =
\frac{1}{(2\pi)^2}\int \mathrm{d}^2 \mathbf{k} \, e^{i\mathbf{k}\cdot
(\mathbf{s}-\mathbf{s}^{\prime\prime})}\widetilde{G}(z,z^{
\prime\prime}
;\mathbf{k};\omega),
\end{equation}
where $\mathbf{r} = (\mathbf{s}, z)$, with $\mathbf{s} = (x,y)$, and 
$\mathbf{k} 
= (k_x,k_y)$. 
The longitudinal Green function corresponding to Eq.~\eqref{eq:ll} is a 
solution of the following equation: 
\begin{equation}
\label{eq:gl}
\left[\nabla_z^2 - m^2\right]\widetilde{G}^{\,\mathrm{L}}(z,z^{
\prime\prime}
;\mathbf{k};\omega)
= \frac{i\lambda^2}{\omega\rho_0}\delta(z-z^{\prime\prime}),
\end{equation}
where $m^2 = \mathbf{k}^2+\lambda^2$ and we 
have defined the longitudinal decay constant $\lambda$ as
\begin{equation}
\lambda^2 = -\frac{i\omega^2\rho_0}{\left(4\eta/3 + \zeta\right)\omega + i 
\rho_0c_0^2}.
\label{eq:lambda}
\end{equation}

The solution of Eq.~\eqref{eq:gl} is well known 
\cite{Erbas,Kim,Schwinger}, and with no-slip boundary conditions at $z = 0$ and 
$z = L$, the Green function is obtained as 
\begin{align}
\widetilde{G}^{\,\mathrm{L}}(z,z^{
\prime\prime}
;\mathbf{k};\omega) = &\, 
{} g_1^{\,\mathrm{L}}e^{-mz}+g_2^{\,\mathrm{L}}e^{m(z-L)} \nonumber\\
& \qquad{} - 
\frac{i\lambda^2 }{2m 
\omega\rho_0}e^{-m|z-z''|},\label{eq:greenl}
\end{align}
where
\begin{align}
g_1^{\,\mathrm{L}} = & {} \frac{i\lambda^2}{2m\omega \rho_0} 
\csch(mL)\sinh(m(L-z'')), \\
g_2^{\,\mathrm{L}} = & {} \frac{i\lambda^2}{2m\omega \rho_0}  \csch(mL)\sinh(m 
z''),
\end{align}
are constants of integration that satisfy the no-slip boundary conditions.

\subsection{Characteristic scales and dimensionless parameters}
\label{subsec:para}

We simplify the following analysis by introducing dimensionless parameters that 
characterize the fluid and the plane-parallel  geometry of our system. There 
are 
two length scales that can be used for this purpose: The macroscopic plate 
separation, $L$, and the microscopic 
scale at which the continuum hydrodynamic description breaks down, which we 
denote $a $. There are two characteristic vorticity frequencies associated 
with each of these length scales \cite{Erbas},
\begin{equation}\label{eq:omega0def}
\omega_0 = \frac{\eta}{L^2\rho_0} \qquad \mathrm{and}\qquad \omega_\infty = 
\frac{\eta}{a ^2\rho_0}.
\end{equation}
The inverse frequencies, $\omega_0^{-1}$ and $\omega_\infty^{-1}$, correspond 
to the time that vorticity requires to diffuse a certain distance, in this case 
$L$ or $a $, respectively. We also define the dimensionless parameter 
$\gamma$, which is given by
\begin{equation}
\gamma = \frac{c_0^2}{L^2\omega_0^2} = \left(\frac{L\rho_0 
c_0}{\eta}\right)^2.\label{eq:gammadef}
\end{equation}
This parameter is the squared ratio of the vorticity time scale and the typical 
compression time scale in which a propagating sound wave travels a distance $L$ 
\cite{Erbas}.

To facilitate our later discussions, we introduce the dimensionless ratios
\begin{equation}\label{eq:udef}
u = \frac{\omega}{\omega_0\gamma} \qquad \mathrm{and} \qquad u_\infty = 
\frac{\omega_\infty}{\omega_0\gamma},
\end{equation}
and define the function
\begin{equation}\label{eq:fudef}
f_m(u) = \frac{u^{2-m}}{1+\chi^2u^2}.
\end{equation}
We can now express the real and imaginary parts 
of the longitudinal decay 
constant, $\lambda$, as
\begin{align}
\ell_+ = {} & \lambda_{\mathrm{R}}L = 
\frac{\omega_0\gamma 
L}{c_0}\frac{|u|}{\sqrt{2}}\sqrt{\left[1-\sqrt{f_2(u)}\right]\sqrt{f_2(u)} } ,\\
\ell_- = {} & \lambda_{\mathrm{I}}L =-
\frac{\omega_0\gamma 
L}{c_0}\frac{u}{\sqrt{2}}\sqrt{\left[1+\sqrt{f_2(u)}\right]\sqrt{f_2(u)} } .
\end{align}

The vorticity frequency scale $\omega_0$ marks the boundary 
between the low-frequency 
propagative regime, for which $\omega < \omega_0\gamma$ (or $u<1$) and sound 
waves 
propagate with speed $c \sim |\lambda_{\mathrm{I}}^{-1}|$, and the 
high-frequency 
diffusive regime, for which $\omega > \omega_0\gamma$ (or $u>1$) and viscosity 
effects damp 
compression perturbations \cite{Erbas}.
Furthermore, the dimensionless ratio $u_\infty$ can be expressed in terms of a 
new length scale $\delta $:
\begin{equation}
\label{eq:u_inf}
u_\infty = \frac{\delta^2}{a ^2} \qquad \mathrm{where}\qquad \delta  = 
\frac{\eta}{\rho_0c_0} = \frac{c_0}{\omega_0 \gamma}.
\end{equation}
This length scale characterizes the boundary between the propagative and 
diffusive regimes at $\omega_0 \gamma$. We can also define a characteristic 
time scale, 
\begin{equation}
t_0 = \delta/c_0, 
\label{eq:t0}
\end{equation}
associated 
with this boundary. Finally, then, we can write $\ell_+$ and $\ell_-$ as 
\begin{align}
\ell_+ = {} & 
\frac{
L}{\delta }\frac{|u|}{\sqrt{2}}\sqrt{\left[1-\sqrt{f_2(u)}\right]\sqrt{f_2(u)} 
} , 
\label{eq:lambdar}\\
\ell_- = {} & -\frac{
L}{\delta }\frac{u}{\sqrt{2}}\sqrt{\left[1+\sqrt{f_2(u)}\right]\sqrt{f_2(u)} 
} . 
\label{eq:lambdai}
\end{align}

For any reasonable choice of realistic parameters for a fluid far from the 
critical point, we have $u\ll1$, i.e., we work in the propagative regime. In 
this case, the plate separation of a realistic experiment satisfies $L/\delta 
\gg~1$. 
For liquids close to the critical point, or polymers in solution, however, the 
crossover frequency can be much lower and, therefore, we can have $u\gg1$. In 
this 
case, the system is in the diffusive regime and the crossover length scale, 
$\delta $, may be macroscopic. 

\subsection{Correlation functions}
\label{sec:corrfns}
 
Now that we have explicit expressions for the Green function solutions in 
hand, we turn to the correlation functions $c_{zz}^{\mathrm{L}\mathrm{L}}$ and
$c^{\rho\rho}$, which enter in Eqs.~\eqref{eq:czzpt}-\eqref{eq:p2t}, and express 
these 
correlation functions in terms of the 
corresponding Green functions. Here, we simply sketch the derivation for 
$c_{zz}^{\mathrm{L}\mathrm{L}}$, as an example, and leave the details of the 
corresponding calculation of $c^{\rho\rho}$ to Appendix
\ref{app:corrfn}.

The longitudinal velocity fluctuations are given in terms of the longitudinal 
Green function as
\begin{equation}\label{eq:vtt}
v_i^{
\mathrm{L}}(\mathbf{r};t) = \int \mathrm{d}t^{\prime\prime}  \int 
\mathrm{d}^3\mathbf{r}^{\prime\prime} 
\,
G^{\,\mathrm{L}}(\mathbf{r},\mathbf{r}^{\prime\prime} ;t-t^{\prime\prime} 
)\Sigma^{
\mathrm{L}}_i(\mathbf{r}^{\prime\prime} ;t^{\prime\prime} ).
\end{equation}
We require the correlation function
\begin{align}\label{eq:vvtrep}
\big\langle  
v_i^{\mathrm{L}}(\mathbf{r};t) & {}
v_j^{\mathrm{L}}(\mathbf{r}^\prime;t') \big\rangle  = \int 
\mathrm{d}t^{\prime\prime} \int 
\mathrm{d}t^{\prime\prime\prime}\int 
\mathrm{d}^3\mathbf{r}^{\prime\prime} \,\int 
\mathrm{d}^3\mathbf{r}^{\prime\prime\prime} \nonumber\\
& {} \times
G^{\,\mathrm{L}}(\mathbf{r},\mathbf{r}^{\prime\prime};t-t^{\prime\prime})G^{\,
\mathrm { L
}}(\mathbf{r}^\prime,\mathbf{r}^{\prime\prime\prime};t'-t^{
\prime\prime\prime } )\nonumber\\
& {} \quad \times\left\langle
\Sigma^{\mathrm{L}}_i(\mathbf{r}^{\prime\prime};t^{\prime\prime})\Sigma^{\mathrm
{ L } } _j(\mathbf {
r}^{\prime\prime\prime};t^{
\prime\prime\prime })\right\rangle.
\end{align}
Recalling the stochastic properties of the random stress tensor, 
Eq.~\eqref{eq:svecLw-2}, we obtain
\begin{align}\label{eq:vvtrep2}
c_{ij}^{\mathrm{L}\mathrm{L}}{} & (\mathbf{r},\mathbf{r}';t,t') 
= 2k_{\mathrm{B}}T\eta\chi  \int 
\mathrm{d}t'' \int 
\mathrm{d}^3\mathbf{r}''  \nonumber\\
& {} \quad \times
\nabla''_iG^{\,\mathrm{L}}(\mathbf{r},\mathbf{r}'';t-t''
)\nabla''_jG^{\,
\mathrm { L
}}(\mathbf{r}',\mathbf{r}'';t'-t'' ).
\end{align}

We now introduce a Fourier representation of the Green functions
\begin{align}\label{eq:cvv}
c_{ij}^{\mathrm{L}\mathrm{L}} {} &(\mathbf{r},\mathbf{r}';t,t') 
= 2k_{\mathrm{B}}T\eta\chi  \int 
\mathrm{d}t'' \int 
\mathrm{d}^3\mathbf{r}'' \nonumber\\
& {} \qquad \times \int\frac{\mathrm{d}\omega}{2\pi} e^{-i \omega (t-t'')}
\nabla''_i\widetilde{G}^{\,\mathrm{L}}(\mathbf{r},\mathbf{r}
'';\omega)\nonumber\\
& {} \qquad \times \int\frac{\mathrm{d}\omega'}{2\pi} e^{-i \omega' 
(t'-t'')}\nabla_j''\widetilde{G}^{\,
\mathrm { L
}}(\mathbf{r}',\mathbf{r}'';\omega' ).
\end{align}
The integral over $t''$ generates a Dirac delta function for the frequencies, 
$\delta(\omega+\omega')$, and therefore one of the frequency integrals is 
trivial:
\begin{align}
c_{ij}^{\mathrm{L}\mathrm{L}} {} &(\mathbf{r},\mathbf{r}';t,t') 
= 2k_{\mathrm{B}}T\eta\chi\int\frac{\mathrm{d}\omega'}{2\pi}e^{i \omega' 
(t-t')} \nonumber\\
& {} \quad \times   \int 
\mathrm{d}^3\mathbf{r}''
\nabla''_i\widetilde{G}^{\,\mathrm{L}}(\mathbf{r},\mathbf{r}
'';-\omega') \nabla_j''\widetilde{G}^{\,
\mathrm { L
}}(\mathbf{r}',\mathbf{r}'';\omega' ).
\end{align}

In principle, we could substitute our explicit expression for the Green 
function, Eq.~\eqref{eq:greenl}, into this 
correlation function and attempt to directly calculate the integrals at this 
stage. We will see, however, that this is not the most straightforward 
approach: Spatial integrations over the fluid boundary will simplify our task 
considerably. We also take advantage of the fact that we only require the 
components of the velocity fields perpendicular to the plane boundaries. 
Therefore, we set $i = j= z$ in our expression for the 
correlation function,
$c_{ij}^{\mathrm{L}\mathrm{L}}(\mathbf{r},\mathbf{r}';t,t')$, and use the 
translational-invariant structure of the Green function, 
Eq.~\eqref{eq:Gdef}, to write
\begin{align}
 c_{zz}^{\mathrm{L}\mathrm{L}} {} &(\mathbf{r},\mathbf{r}';t,t') =
2k_{\mathrm{B}}T\eta\chi 
\int\frac{\mathrm{d}\omega'}{2\pi}e^{i 
\omega' 
(t-t')} \nonumber\\
& \times \int 
\mathrm{d}^3\mathbf{r}''\int \frac{\mathrm{d}^2 
\mathbf{k}}{(2\pi)^2}
e^{i\mathbf{k}\cdot
(\mathbf{s}-\mathbf{s}'')}\nabla''_z 
\widetilde{G}^{\,\mathrm{L}}(z,z'' 
;\mathbf{k};-\omega') \nonumber\\
& \times \int \frac{\mathrm{d}^2 
\mathbf{k}'}{(2\pi)^2}\,e^{i\mathbf{k}'\cdot
(\mathbf{s}'-\mathbf{s}'')} \nabla''_z 
\widetilde{G}^{\,\mathrm{L}}(z',
z'';\mathbf{k}';\omega').
\end{align}

The double integral over $\mathbf{s}''$ generates a 
wavenumber Dirac delta function, $\delta (\mathbf{k}+\mathbf{k}')$, that 
enables us to carry out one of the wavenumber integrals immediately and, thus, 
obtain
\begin{align}
 c_{zz}^{\mathrm{L}\mathrm{L}} {} &(\mathbf{r},\mathbf{r}';t,t') =
2k_{\mathrm{B}}T\eta\chi
\int\frac{\mathrm{d}\omega'}{2\pi}e^{i\omega' (t-t')} \nonumber\\
& \times \int 
\mathrm{d}z''\int \frac{\mathrm{d}^2 
\mathbf{k}}{(2\pi)^2}
e^{i\mathbf{k}\cdot
(\mathbf{s}-\mathbf{s}')}\nabla''_z 
\widetilde{G}^{\,\mathrm{L}}(z,z'' 
;\mathbf{k};-\omega') \nonumber\\
& \qquad \times  \nabla''_z 
\widetilde{G}^{\,\mathrm{L}}(z',
z'';-\mathbf{k};\omega')
.\label{eq:ttcorr}
\end{align}

Analogous arguments apply to the density
correlation function, which is (see Appendix \ref{app:corrfn})
\begin{align}
{} & c^{\rho\rho}(\mathbf{r},\mathbf{r}';t,t') =  \frac{k_{\mathrm{B}}T}{\pi} 
\rho_0^2 \eta\chi 
\int \frac{\mathrm{d}\omega'}{\omega'^2} e^{i\omega'(t-t')}\int 
\mathrm{d}z'' \nonumber\\
{} & \quad \times \int \frac{\mathrm{d}^2 
\mathbf{k}}{(2\pi)^2}\,e^{i\mathbf{k}\cdot
(\mathbf{s}-\mathbf{s}')}\left(\nabla_z 
\nabla''_z+\mathbf{k}^2\right)  
\widetilde{G}^{\,\mathrm{L}}(z,z''; \mathbf{k} ;-\omega') \nonumber \\
{} & \qquad \times \left(\nabla_z' 
\nabla''_z+\mathbf{k}^2\right)  
\widetilde{G}^{\,\mathrm{L}}(z',z''; -\mathbf{k} ;\omega').\label{eq:ppcorr}
\end{align}

\subsection{Spatial integration over surface boundaries}
\label{sec:singleforce}

Our final step is to integrate the correlation functions, 
Eqs.~\eqref{eq:ttcorr} and \eqref{eq:ppcorr}, over the 
boundaries of the fluid according to Eqs.~\eqref{eq:czzpt}-\eqref{eq:p2t}. 
These integrals 
give our final result for the time-dependent correlators of the force 
acting on the fluid boundaries.

The double integrals over $(x,y)$ and $(x^\prime,y^\prime)$ in 
Eqs.~\eqref{eq:czzpt}-\eqref{eq:p2t} lead to a Dirac delta function over the 
transverse wavenumbers, $(2\pi)^2 A \delta(\mathbf{k})$. Thus, we can write 
these equations in terms of 
the Green function as
\begin{align}
& {\cal P}_{1}(z, z'; t,t') =\frac{k_{\mathrm{B}}T}{\pi} \eta^3\chi^3 A 
\int
\mathrm{d}\omega'\cos[\omega'(t-t')]
\nonumber \\
& \times \int
\mathrm{d}z''
\nabla_z
\nabla^{
\prime\prime}_z \widetilde{G}^{\,\mathrm{L}}(z,z^{
\prime\prime}
;\mathbf{0};-\omega')\nabla_z' \nabla''_z \widetilde{G}^{\,\mathrm{L}}(z',z''
;\mathbf{0};\omega'),\label{eq:p1zero}\\
& {\cal P}_2(z, z'; t,t') =\frac{k_{\mathrm{B}}T}{\pi} \rho_0^2\eta\chi c_0^4 
A\int
\frac{\mathrm{d}\omega'}{\omega^{\prime\,2}}\cos[\omega'(t-t')] 
 \nonumber\\
& \times \int
\mathrm{d}z''
 \nabla_z
\nabla^{
\prime\prime}_z \widetilde{G}^{\,\mathrm{L}}(z,z^{
\prime\prime}
;\mathbf{0};-\omega')\nabla_z' \nabla''_z \widetilde{G}^{\,\mathrm{L}}(z',z''
;\mathbf{0};\omega').\label{eq:p2zero}
\end{align}
These frequency integrals run over the frequency range
$\omega \in [-\omega_\infty,\omega_\infty]$ and the spatial integral is over 
$z\in [0,L]$. In writing the above relations, we have used the fact that the 
integrands involved in calculating ${\cal P}_1$ and ${\cal P}_2$ (see 
Eqs.~\eqref{eq:ttcorr} and \eqref{eq:ppcorr}) have odd imaginary parts, which 
thus 
vanish, leading to the factor $\cos[\omega'(t-t')]$ from the real part of the 
exponential factor $e^{i\omega'(t-t')}$. We also 
note that $\widetilde{G}^{\,\mathrm{L}\,\ast}(z',z'';\mathbf{0};\omega') = 
\widetilde{G}^{\,\mathrm{L}}(z',z'';\mathbf{0};-\omega')$, which follows from 
the reality of $G^{\,\mathrm{L}}(z',z'';\mathbf{0};t)$. Therefore, as 
expected, the final correlators are purely real.

Carrying out the derivatives and the remaining spatial integral is
fairly straightforward. The results are 
\begin{align}
{\cal P}_{1}(z, z^\prime; t,t') = {} & \frac{k_{\mathrm{B}}T}{\pi} 
\frac{\rho_0 c_0^2\,  A }{L}\chi^3 \Big[{\cal W}_0(z,z';\tau)\nonumber\\
{} & \quad\qquad  +L {\cal V}_0(z,z';\tau) \delta 
(z-z')\Big],\label{eq:p1zero123}\\
{\cal P}_2(z, z^\prime; t,t') = {} &  \frac{k_{\mathrm{B}}T}{\pi} 
\frac{\rho_0 
c_0^2\, 
 A }{L}\chi\Big[{\cal W}_2(z,z';\tau)\nonumber\\
{} & \quad\qquad+
L {\cal V}_2(z,z';\tau) \delta (z-z')\Big].\label{eq:p2zero123}
\end{align}
The relevant frequency integrals are given by  (see 
Appendix \ref{app:wmderiv})
\begin{align}
{} &{\cal W}_m (0,0;\tau)
=  2\int_0^{u_\infty} 
\mathrm{d}u \, f_m(u)\cos[u\tau] \nonumber\\
{} & \quad \times \frac{1}{\cosh[2\ell_+]
-\cos[2\ell_-]} 
\bigg[\left(\frac{\ell^2}{2\ell_-} 
-2\ell_-\right) \sin[2\ell_-]
\nonumber\\
{} &\qquad \qquad  +\left(\frac{\ell^2}{2\ell_+} 
-2\ell_+\right) \sinh[2\ell_+]\bigg],\label{eq:wmt}\\
{} &{\cal W}_m (0,L;\tau)   = 2\int_0^{u_\infty} 
\mathrm{d}u\,f_m(u)\cos[u\tau] \nonumber\\
{} & \quad\times \frac{1}{\cosh[2\ell_+]
-\cos[2\ell_-]} 
\bigg[\bigg(\frac{\ell^2}{\ell_-} - 4\ell_-\bigg)\cosh[\ell_+]
\sin[\ell_-] \nonumber\\
{} & \qquad\qquad  + \bigg(\frac{\ell^2}{\ell_+} - 4\ell_+\bigg)\cos[\ell_-]
\sinh[\ell_+]\bigg],
\label{eq:wmcrosst} 
\end{align}
and
\begin{equation}
{\cal V}_m(0,0;\tau) = 
                     2{\displaystyle \int_0^{u_\infty} }
\mathrm{d}u \, f_m(u)\cos[u\tau].
\label{eq:vmt0}
\end{equation}
In these equations $\ell^2 = \ell_+^2+\ell_-^2$, where we have defined $\ell_+$ 
and 
$\ell_-$ in Eqs.~\eqref{eq:lambdar} and \eqref{eq:lambdai} and the function 
$f_m(u)$ in Eq.~\eqref{eq:fudef}. The 
dimensionless time parameter is
\begin{equation}
\tau = (t-t')/t_0, 
\end{equation}
with $t_0$ the 
characteristic microscopic timescale defined in Eq.~\eqref{eq:t0}. We have used 
the 
symmetry of the integrand  
to integrate over the positive real axis up to the dimensionless 
microscopic cutoff, $u_\infty$, of Eqs.~\eqref{eq:udef} and \eqref{eq:u_inf}. 
To simplify these expressions further, we note that
\begin{equation}
{\cal V}_2(0,0;\tau)+\chi^2{\cal V}_0(0,0;\tau) 
= \frac{2}{\tau}\sin[u_\infty \tau].\label{eq:vmt}
\end{equation}

Putting together all of these results, from Eqs.~\eqref{eq:czzpt} and 
 \eqref{eq:p1zero}-\eqref{eq:vmt}, we find
 \begin{align}
 {\cal C}(0,0;t,t')
=  {} &   \frac{k_{\mathrm{B}}T}{\pi}    
\frac{\rho_0 c_0^2  A}{L} \chi\bigg[\frac{2L}{\tau}\sin[u_\infty \tau]\delta(0) 
 \nonumber\\
 {} & \quad +\chi^2 {\cal W}_0(0,0;\tau)
 + {\cal 
W}_2(0,0;\tau) \bigg],
\label{eq:final1}
\\
 {\cal C}(0,L;t,t')
= {}&  \frac{k_{\mathrm{B}}T}{\pi}    
\frac{\rho_0 c_0^2  A}{L} \chi \nonumber\\
{} & \quad \times \Big[ \chi^2
 {\cal W}_0(0,L;\tau) + {\cal 
W}_2(0,L;\tau)\Big].
\label{eq:final2}
\end{align}

These are our final results: The same-plate and the 
cross-plate correlators of the normal force, expressed in terms of the four 
frequency 
integrals ${\cal W}_m(0,0;\tau)$ and ${\cal W}_m(0,L;\tau)$ where $m=0, 2$. 
Thus, while the average fluctuation-induced force between the bounding surfaces 
vanishes identically (see Sec.~\ref{sec:meanforce}), the correlation 
functions of the force show a pronounced dependence on the inter-plate 
separation and the time difference.

The same-plate correlator of the normal force, $ {\cal C}(0,0;t,t')$ 
in Eq.~\eqref{eq:final1}, contains three terms.
The first contribution is local in space and therefore proportional to the Dirac 
delta 
function. Comparing this first term to ${\cal 
P}_0(z,z'; t, t')$ in Eq.~\eqref{eq:p0delta} indicates
that this contribution to $ {\cal C}(0,0;t,t')$ is related to the integral of 
the random stress correlator, Eq.~\eqref{eq:gaussprop2}, across 
the bounding surfaces, but with the hydrodynamic coupling nevertheless
fully taken into account. Incorporating the hydrodynamic coupling leads to
non-locality in time, while 
locality in space is preserved at leading order. 
In contrast, ${\cal P}_0(z,z'; t, t')$ is local both in 
time and in space, because this term follows directly from the correlator of 
the random stress tensor without any hydrodynamic coupling and has, 
therefore, been dropped from our present analysis. 
The other two terms in $ {\cal C}(0,0;t,t')$ are different in nature. They are 
non-local both in time 
and in space. They correspond to self-correlations mediated by the hydrodynamic 
interaction between the boundaries, leading to separation-dependent 
contributions to the same-plate, normal force correlator. 
These two terms present a non-trivial generalization of the normal force 
correlator that hydrodynamically
couples the boundaries. We now define these contributions to be the {\em 
excess} correlator,
\begin{align}
\Delta{\cal C} (0,0; t, t')
\equiv  {} & \frac{k_{\mathrm{B}}T}{\pi}   \frac{\rho_0 c_0^2\,  A }{L} \chi 
\nonumber\\
{} & \times 
\bigg[\chi^2
 {\cal W}_0(0,0; \tau) + {\cal W}_2(0,0; \tau)\bigg],
 \label{eq:DeltaC_t}
\end{align}
which we investigate in detail in the following sections. 

The cross-plate correlator of the normal force in Eq.~\eqref{eq:final2} 
does not contain any local terms. In fact, it is purely
non-local and does not include any hydrodynamic self-interactions. The 
cross-plate correlator is due entirely to hydrodynamic interactions 
across the fluid between the boundaries, and thus 
naturally depends on the boundary separation. 

In summary, for the same-plate force correlator, we have identified a trivial 
term that is local in space and non-trivial terms that are non-local in space 
and correspond to 
self-correlations mediated by the hydrodynamic coupling between the 
boundaries. This leads to the separation-dependent excess same-plate force 
correlator. On the 
other hand, the cross-plate force correlator contains no local terms, as 
expected, and 
stems entirely from hydrodynamic interactions between the bounding surfaces.

\section{Results for equal-time force correlators}
\label{sec:numerics}

\begin{table}[t!]
\caption{\label{tab:parms}Representative ranges of physical parameters in a 
realistic fluid; see the text for definitions.
\\}
\begin{ruledtabular}
\begin{tabular}{ccc}
\vspace*{-5pt}\\
Parameter & Description & Range \\
\vspace*{-5pt}\\
\hline
\vspace*{-5pt}\\
$L $ & Plate separation &  10$^{-6}$ to 10$^{-3}$ m \\
$\delta $ & Propagative-diffusive boundary & 10$^{-9}$ to 10$^{-6}$ m \\
$a $ & Microscopic cutoff & 10$^{-9}$ m \\
$\eta$ & Shear viscosity & 10$^{-4}$ to 1 Pa$\cdot$s \\
$\zeta$ & Bulk viscosity & 10$^{-4}$ to 1 Pa$\cdot$s \\
\end{tabular}
\end{ruledtabular}
\end{table}

Our task now is to explore and evaluate the frequency integrals that appear in 
Eqs.~\eqref{eq:final1}-\eqref{eq:DeltaC_t}. We start by considering the 
equal-time correlators that follow from these equations by setting $\tau=0$ or, 
equivalently,  $t=t'$, i.e., 
\begin{align}
\Delta{\cal C} (0,0)
=  {}&  \frac{k_{\mathrm{B}}T}{\pi}   \frac{\rho_0 c_0^2\,  A }{L} \chi 
\bigg[\chi^2
 {\cal W}_0(0,0) + {\cal W}_2(0,0)\bigg],
 \label{eq:DeltaC} \\
{\cal C}(0,L) = {} & \frac{k_{\mathrm{B}}T}{\pi}   
\frac{\rho_0 c_0^2\,  A }{L} \chi 
\bigg[ \chi^2
 {\cal W}_0(0,L) + {\cal W}_2(0,L)\bigg],\label{eq:ffp5678}
\end{align}
where we have set ${\cal C}(z,z')\equiv {\cal C}(z,z';t,t) $ and $ {\cal 
W}_m(z,z')\equiv {\cal W}_m(z,z';\tau=0)$. We note that $\Delta{\cal C} (0,0)$ 
is, in fact, the excess \emph{force variance}.

The dimensionless integrals, ${\cal W}_m(z,z')$, are functions of 
just three dimensionless ratios: The ratio of the fluid viscosities, 
$\zeta/\eta$, which enters through $\chi$, defined in Eq.~\eqref{eq:chi}; the 
ratio of the plate separation to the propagative-diffusive boundary length 
scale $L/\delta $; and the ratio of the propagative-diffusive boundary length 
scale to the microscopic cutoff scale, $u_\infty = (\delta /a )^2$. 

\begin{figure}[t!]
\centering
\includegraphics[width=7.5cm]{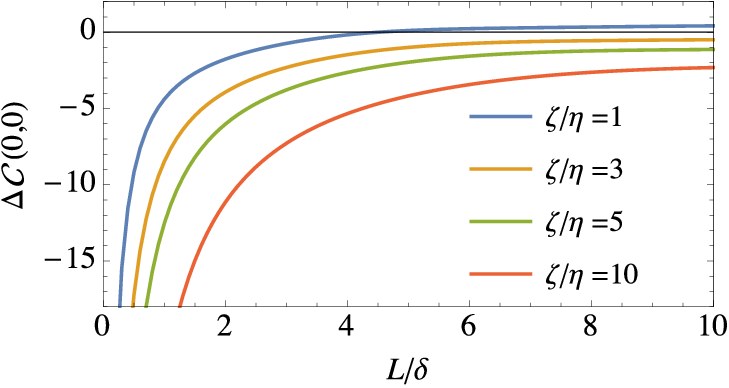} (a)
\\
\vspace{.3cm}
\includegraphics[width=7.5cm]{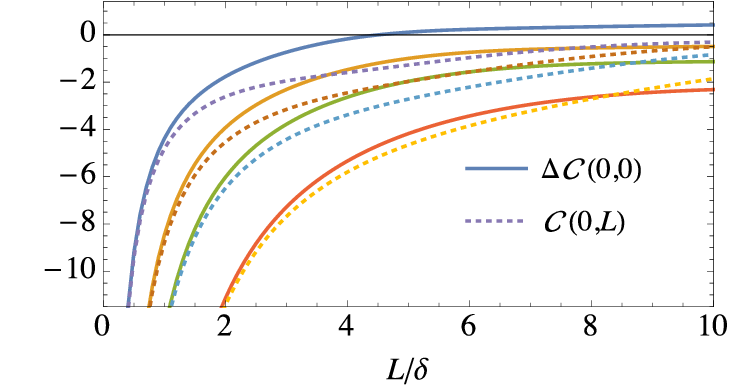} (b)
\\
\vspace{.3cm}
\includegraphics[width=7.5cm]{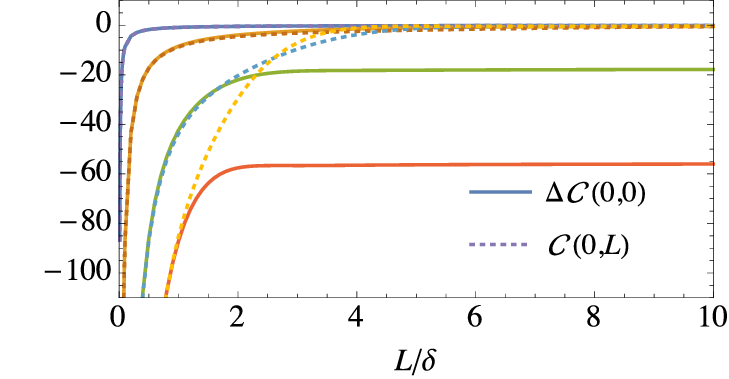} (c)
\caption{\label{fig:fig1} 
(Color online) (a) Equal-time, excess same-plate force correlator 
(or force variance), $\Delta {\cal C}(0,0)$, as defined in 
Eq.~\eqref{eq:DeltaC}, plotted as a function of $L/\delta$ for fixed  $u_\infty 
= (\delta/a)^2 = 1$ and 
$\zeta/\eta=1, 3, 5$ and 10 (top to bottom). (b) Same as (a) but here we plot 
the equal-time, cross-plate force correlator ${\cal C}(0,L)$ as defined in 
Eq.~\eqref{eq:ffp5678} (dashed curves) and compare it with the equal-time, 
excess 
same-plate force correlator (solid curves). (c) Comparison of equal-time, 
excess same-plate force correlator, $\Delta {\cal C}(0,0)$ (solid curves), and 
equal-time, cross-plate force correlator, ${\cal C}(0,L)$ (dashed curves), 
plotted as a function of $L/\delta$ for fixed $\zeta/\eta=3$ and $u_\infty=0.1, 
1.0, 5.0$ and 10.0 (top to bottom). We plot the force correlators in 
units of 
$(k_{\mathrm{B}}T/\pi)\cdot(\rho_0 c_0^2\, A)$.  
}
\end{figure}

We tabulate our choices for these parameters, which correspond to a range of 
reasonable physical values, in Table \ref{tab:parms}. For the 
case of confined fluids, relevant in particular for our analysis, 
experiments and simulations on nano-slit confined water suggest that the 
bulk viscosity is recovered at boundary surface separations larger than 
approximately one nanometer \cite{Raviv1,Raviv2,Leng}. To be on the 
safe side, we therefore take one nanometer as the implied microscopic cutoff 
$a$, but also indicate in Fig.~\ref{fig:fig1}(c) how the equal-time force 
correlators  depend on this 
cutoff through the dimensionless parameter $u_\infty$.

We evaluate the frequency integrals $ {\cal W}_m(z,z')$ numerically and plot 
the excess force variance, $\Delta{\cal C} (0,0)$, as a function 
of $L/\delta$ for $\zeta/\eta=$1, 3, 5 and 10 in Figs.~\ref{fig:fig1}(a) and 
\ref{fig:fig1}(b) (solid curves). The cross-plate correlator, ${\cal C} (0,L)$, 
is shown by dashed curves in Fig.~\ref{fig:fig1}(b), where, for the sake of 
comparison, the curves for $\Delta{\cal C} (0,0)$ are replotted. In the 
figures, 
we plot the force correlators in units of $(k_{\mathrm{B}}T/\pi)\cdot(\rho_0 
c_0^2\, A)$.  

Figs.~\ref{fig:fig1}(a) and \ref{fig:fig1}(b) show that both $\Delta{\cal C} 
(0,0)$ and ${\cal C} (0,L)$ become negative at 
{\em small separations}, $L/\delta\ll 1$, and eventually diverge when 
$L/\delta\rightarrow 0$. In this limit, the curves for both these correlators 
overlap and, thus, they are approximated by the same limiting form. At {\em 
large separations}, $L/\delta\gg 1$, the cross-plate correlator tends to zero 
while the excess same-plate correlator tends to a constant depending on the 
viscosity parameters.
Therefore, the cross-plate correlator remains negative over the whole range of 
separations, indicating that the two bounding surfaces are subjected to {\em 
counter-phase correlations}. The excess same-plate correlator, on the other 
hand,  can be negative (for intermediate to large values of $\zeta/\eta$) or 
positive (for sufficiently small $\zeta/\eta$). Thus, when the two correlators 
are compared, as in Fig.~\ref{fig:fig1}(b), one can see that, at small to 
intermediate separations, the cross-plate correlator (dashed curves) is  
larger in magnitude than the excess same-plate correlator (solid curves); 
while, 
at large separations, it can become smaller than the latter. The difference 
between these two quantities decreases with increasing $\zeta/\eta$. 

In addition, Fig.~\ref{fig:fig1}(c) demonstrates that the two correlators 
overlap for the whole range of plate separations for small $u_\infty$, 
illustrated by the overlap of the solid and dashed blue curves at 
$u_\infty=0.1$. For large $u_\infty$, such as at $u_\infty=10.0$, indicated by 
the solid red and dashed yellow curves, these correlators deviate significantly. 
At large plate separations, the same-plate correlator tends to a value that 
is independent of the plate separation, in agreement with the analytic result 
of Eq.~\eqref{eq:c00si}, while the cross-plate correlator becomes independent 
of $u_\infty$ and decays with the separation, in agreement with 
Eq.~\eqref{eq:largeL}. 

\subsection{Analytic limits}

\begin{figure}[t]
\centering
\includegraphics[width=7.5cm]{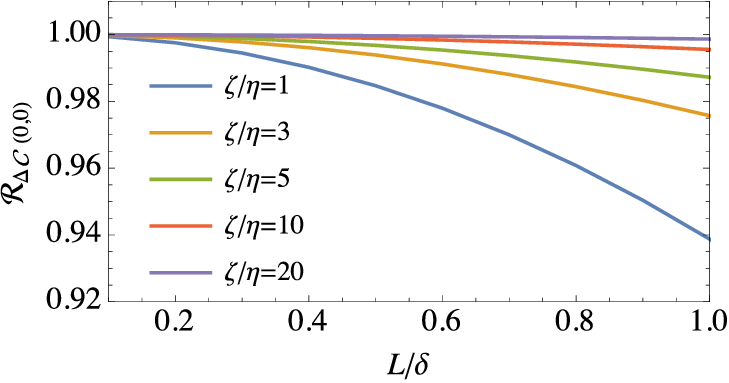} (a)
\\
\vspace{.3cm}
\includegraphics[width=7.5cm]{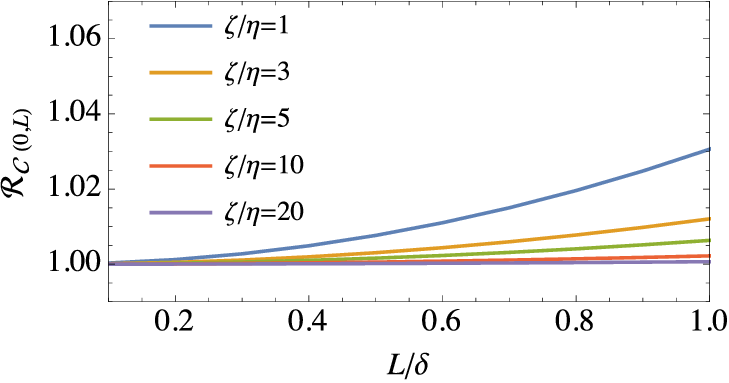} (b)
\caption{\label{fig:c0LsmallL} 
(Color online) (a) Ratio of full numerical results of $\Delta{\cal C}(0,0)$ to 
the analytic limiting behavior of Eq.~\eqref{eq:smallL}, shown on the 
graph by ${\cal R}_{{\Delta C}(0,0)}$, as a
function of $L/\delta $ for $L/\delta \leq 1$ and $\zeta/\eta=1, 3, 5, 10$ and 
20. We fix 
$u_\infty = (\delta /a )^2 = 1$. (b) Same as (a) but for ${\cal 
C}(0,L)$. }
\end{figure}

Beyond these numerical results, we can analytically calculate the small and 
large plate-separation limits and the limits of vanishing and infinite 
speed of sound (Burger's and incompressible limits, respectively). To 
study 
the small and large
plate-separation cases, we first note that the 
frequency integrals of Eqs.~\eqref{eq:wmt} and \eqref{eq:wmcrosst} depend on 
$L/\delta $ through $\ell_+$ and $\ell_-$, which are both linear in this ratio 
(see Eqs.~\eqref{eq:lambdar} and \eqref{eq:lambdai}).

Thus, in the {\em small separation limit}, $L/\delta \ll 1$, we can expand the 
frequency integrands as Taylor series in $L/\delta $ for both ${\cal W}_m 
(0,0)$ 
and ${\cal W}_m (0,L)$ and keep terms up to linear order in $L/\delta $. 
The resulting integrals are trivial, giving 
\begin{equation}
\Delta{\cal C} (0,0)
\,{\simeq }\,  {\cal C} (0,L) 
\,{\simeq }\, 
-\frac{k_{\mathrm{B}}T}{\pi} 
\frac{2 \left(4\eta/3+\zeta\right)\eta  A}{\rho_0 a^2 L}.\label{eq:smallL}
\end{equation}
These expressions agree with the full numerical results for $L/\delta \ll 1$, as
we illustrate in
Fig.~\ref{fig:c0LsmallL}. This figure shows the ratio of the full numerical 
result to the analytic approximation of 
Eq.~\eqref{eq:smallL} for  $\Delta{\cal 
C}(0,0)$ (panel a) and ${\cal C}(0,L)$ (panel b). 
The plots show that the ratio in both cases tends to unity as $L/\delta$ 
becomes sufficiently small, but the domain of validity of the 
analytic approximation depends strongly on the ratio $\zeta/\eta$ and 
increases with increasing $\zeta/\eta$.

In the {\em large separation limit}, $L/\delta\gg 1$, the 
force variance reduces to the semi-infinite fluid result (see Appendix 
\ref{app:othergeom}),
\begin{align}\label{eq:c00si}
{} & \Delta{\cal C}(0, 0) \stackrel{L/\delta\rightarrow \infty}{=}
\frac{k_{\mathrm{B}}T}{\pi} 
\frac{2\rho_0^2c_0^3A}{\eta\chi}\bigg[\frac{
2\sqrt { z_\infty-1 } } { z_\infty}\nonumber\\
{} & \quad +\frac{8\sqrt{2}}{3}
\frac{z_\infty(3-z_\infty)}{\sqrt{z_\infty-1}}\sin^4\left(\frac{1}{2}
\arctan(z_\infty^2-1)\right)\bigg],
\end{align}
where $z_\infty=\sqrt{1+x_\infty^2}$ and $x_\infty = \chi u_\infty = 
\chi\eta^2/(a^2\rho_0^2c_0^2)$.

\begin{figure}[t]
\centering
\includegraphics[width=7.5cm]{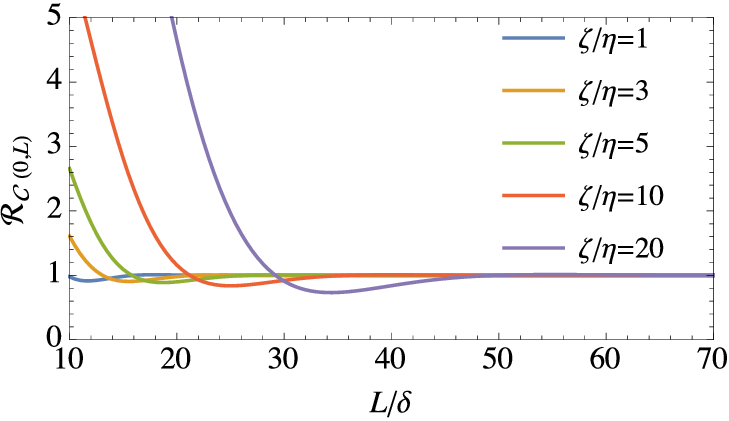}
\caption{\label{fig:c0LlargeL} 
(Color online) Ratio of full numerical results of ${\cal C}(0,L)$ to the 
analytic limiting behavior of Eq.~\eqref{eq:largeL}, shown on the graph by 
${\cal R}_{C(0,L)}$, as a
function of $L/\delta $ for $L/\delta \gg 1$ and $\zeta/\eta=1, 3, 5, 10$ and 
20. We fix 
$u_\infty = (\delta /a )^2 = 1$.}
\end{figure}

The corresponding equal-time, cross-plate  correlator 
tends to zero in the large
plate-separation limit as 
\begin{equation}
{\cal C}(0,L)
\,{\simeq}\,
-\frac{k_{\mathrm{B}}T}{\pi} 
\frac{\pi\rho_0c_0^2  A}{L}.\label{eq:largeL}
\end{equation}
This limiting behavior is independent of the ratio $\zeta/\eta$, as is clearly 
demonstrated by the plots of the ratio of the full numerical result to the 
analytic approximation of 
Eq.~\eqref{eq:largeL} in Fig.~\ref{fig:c0LlargeL}. However, the exact value 
of $L/\delta$ beyond which 
Eq.~\eqref{eq:largeL} is a reasonable approximation does depend on $\zeta/\eta$.

In the {\em incompressible fluid limit}, we consider the 
leading contributions for $c_0\rightarrow \infty$, giving
\begin{equation}
\Delta{\cal C} (0,0) \stackrel{c_0\rightarrow 
\infty}{=} {\cal C} (0,L) \stackrel{c_0\rightarrow 
\infty}{=}  
-\frac{k_{\mathrm{B}}T}{\pi} 
\frac{2 \left(4\eta/3+\zeta\right)\eta  A}{\rho_0 a^2 L}.\label{eq:c0infty}
\end{equation}
There is a correspondence between the incompressible fluid limit and the small 
plate-separation limiting result 
of Eq.~\eqref{eq:smallL}: At small separations, the fluid behaves as if it were 
incompressible. 

In the limit of vanishing adiabatic speed of sound (``Burger's limit''), 
$c_0\rightarrow 0$, 
on the other hand, 
both $C(0,0)$ and $C(0,L)$ tend to zero as $c_0^2$.

\section{Results for time-dependent correlators}
\label{sec:tnumerics}

We now turn to the two-point, time-dependent correlators of the normal forces 
acting on
the walls, which we compute numerically using Eqs.~\eqref{eq:final1} and 
\eqref{eq:final2} for the same-plate and the cross-plate correlators, ${\cal  
C}(0,0;t,t')$ and ${\cal  C}(0,L;t,t')$, respectively. 

\begin{figure}[t!]
\centering
\includegraphics[width=7.5cm]{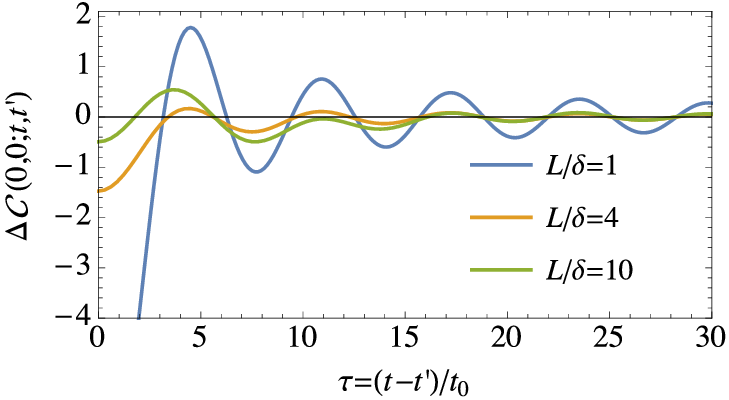}(a)
\\
\vspace{.3cm}
\includegraphics[width=7.5cm]{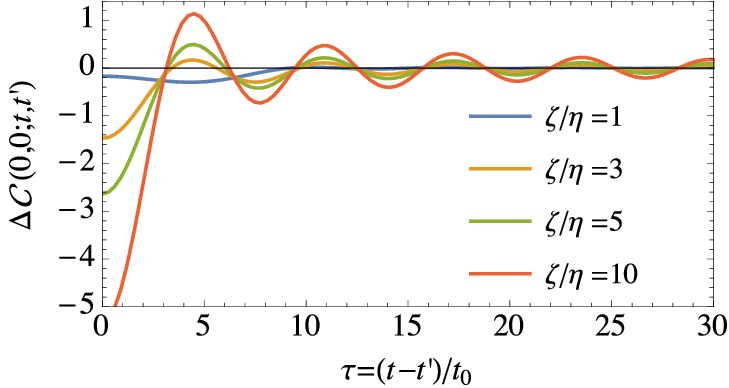}(b)
\caption{\label{fig:fig2}
(Color online) (a) Time-dependent, excess same-plate force correlator,
$\Delta {\cal C}(0,0;t, t')$, as defined in Eq.~\eqref{eq:final1}, plotted as 
a function of the rescaled time difference, $\tau=(t-t')/t_0$, for fixed  
$u_\infty = (\delta/a)^2 = 1$, $\zeta/\eta=3$ and $L/\delta=1, 4$ and 10, as 
indicated on the graph. (b) Same as (a) but here we show the results for fixed 
$u_\infty = (\delta/a)^2 = 1$, $L/\delta=4$ and $\zeta/\eta=1, 3, 5$ and 10.  
}
\end{figure}

We plot the behavior of the excess force correlator, Eq.~\eqref{eq:DeltaC_t}, 
as a function of the rescaled time difference, $\tau=(t-t')/t_0$, for
rescaled inter-plate separations $L/\delta=1, 4$ and 10 in 
Fig.~\ref{fig:fig2}(a). In Fig.~\ref{fig:fig2}(b), we show the time-dependent 
behavior of the same quantity for $\zeta/\eta=1, 3, 5$ and 10.

\begin{figure}[t!]
\centering
\includegraphics[width=7.5cm]{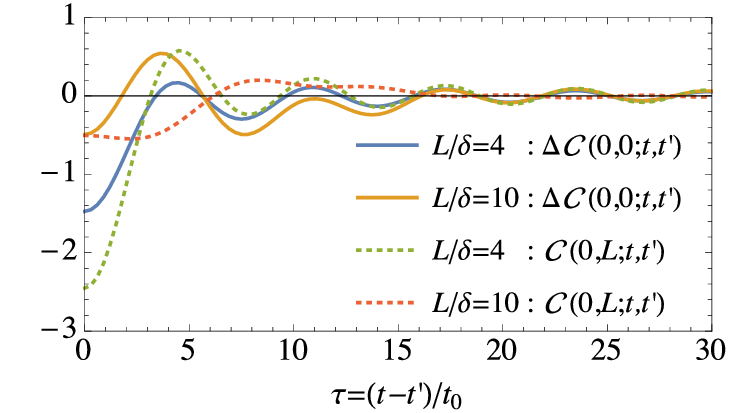}
\caption{\label{fig:c00t}
(Color online) Time-dependent, excess same-plate force correlator, 
$\Delta {\cal C}(0,0;t, t')$ (solid curves), compared with the 
time-dependent cross-plate correlator, ${\cal C}(0,L;t, t')$ (dashed curves), 
for fixed  $u_\infty = (\delta/a)^2 = 1$, $\zeta/\eta=3$ and at two different 
rescaled inter-plate separations, $L/\delta=4$ and 10, as indicated on the 
graph. 
}
\end{figure}

As seen in these figures, $\Delta {\cal C}(0,0;t, t')$ exhibits a 
damped {\em oscillatory} behavior in $\tau$. For $L/\delta \lesssim 3$, these 
oscillations are well  described by a function of the form 
$\alpha\sin(u_\infty\tau)/\tau$, where 
$\alpha$ is a function of the viscosity ratio, $\zeta/\eta$, the 
dimensionless 
cutoff, $u_\infty$, and the rescaled plate separation, $L/\delta$. For 
$L/\delta 
\gtrsim 3$, this simple behavior breaks down, although $\Delta {\cal C}(0,0;t, 
t')$ remains oscillatory with an amplitude that gradually decreases for large 
$\tau$. For the example of water at room 
temperature, with the plate separation $L/\delta = 1$, $\zeta/\eta = 3$ 
and cutoff $u_\infty 
= 1$, we find $\alpha = -8.5(3)$. 

The cross-plate force correlator shows a similar 
time-dependent behavior as the excess same-plate correlator, and the onset 
of irregular oscillations occurs for similar values of $L/\delta$. We
compare the same-plate (solid curves) and cross-plate (dashed curves) 
correlators in Fig.~\ref{fig:c00t}. 

\begin{figure}[t!]
\centering
\includegraphics[width=7.5cm]{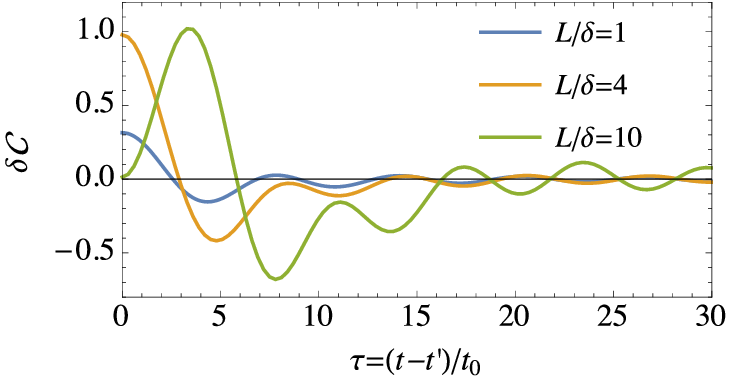}(a)
\\
\vspace{.3cm}
\includegraphics[width=7.5cm]{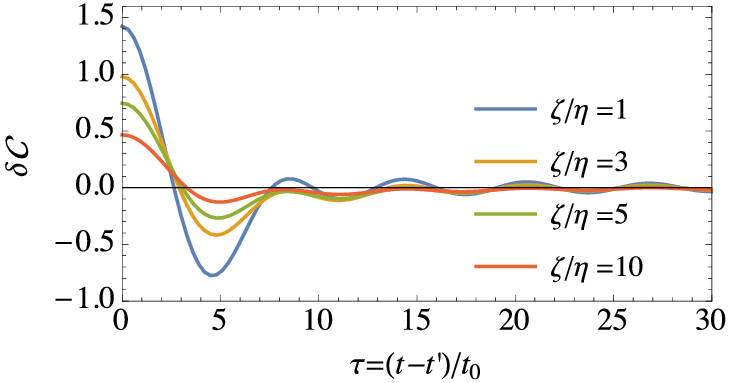}(b)
\caption{\label{fig:fig4} 
(Color online) (a) The difference between the
excess same-plate and the cross-plate force correlators defined as $\delta 
{\cal C} \equiv \Delta {\cal C}(0,0;t, t')-{\cal C}(0,L;t, t')$, plotted as 
a function of the rescaled time difference, $\tau=(t-t')/t_0$, for fixed  
$u_\infty = (\delta/a)^2 = 1$ and $\zeta/\eta=3$, and $L/\delta=1, 4$ 
and 10. (b) Fixed $L/\delta=4$ and $\zeta/\eta=1, 3, 5$ and 10 as indicated 
on the graph. 
}
\end{figure}

We plot the difference between the excess same-plate correlator and the 
cross-plate 
correlator, defined as $\delta {\cal C} \equiv \Delta {\cal C}(0,0;t, t')-{\cal 
C}(0,L;t, t')$, in Fig.~\ref{fig:fig4}. This plot shows that 
the two correlators exhibit similar period of oscillations for a wide range of 
viscosities, especially at small to intermediate  inter-plate separations. In 
the special case of equal-time correlators with $\tau=0$, one can see a 
non-monotonic behavior for the two correlators in Fig.~\ref{fig:fig4}(a): At 
small inter-plate 
separations, $\delta {\cal C}|_{\tau=0}$ is positive and increases by 
increasing $L/\delta$, but this trend changes at around $L/\delta\simeq 3$, and 
then tends to zero for large $L/\delta$.

\section{Conclusion and Discussion}
\label{sec:conclusions}

We have revisited the problem of long-range, fluctuation-induced 
(or  Casimir-like) hydrodynamic interactions within the context of 
Landau-Lifshitz's 
linear, stochastic hydrodynamics in a classical, compressible, viscous fluid 
confined between two rigid, planar walls with no-slip boundary conditions and 
in the absence of heat transfer. We 
show conclusively that, at this level 
and within the pertinent approximations, there is {\em no standard or primary 
Casimir effect} manifest in the average value of the interaction 
force between the fluid boundaries. Nevertheless, we show that there does exist 
a {\em secondary Casimir 
effect} in the {\em variance of the normal force} as well as in the 
{\em cross-correlation function} of the normal force between the bounding 
surfaces. Fluctuations in such effective fluctuation-induced forces have been 
investigated in other Casimir-like contexts \cite{Bartolo, Dean} 
and in disordered charged systems \cite{disorder-PRL,pre2011,epje2012}. 

We derive general expressions for  the two-point, time-dependent, force 
correlations and, thus, show that: 
\begin{enumerate}
\item The variance of the fluctuation-induced force is finite and 
independent of the separation between the bounding surfaces for large 
separations; 
\item The equal-time, cross-plate force correlation exhibits a 
long-range decay with the inverse plate separation that is independent of the 
fluid viscosities;
\item The time-dependent force correlations exhibit a damped oscillatory 
behavior for 
small and intermediate inter-plate separations that grows more irregular at 
large separations.
\end{enumerate}

Our calculation is based on the Landau-Lifshitz linear stochastic hydrodynamics 
and, therefore, does not include putative non-linear effects \cite{Jones}. If 
such 
effects could be brought into the fold, they would have to be considered 
consistently for all variables. Moreover, we find that incorporating 
compressibility 
does not completely obliterate all fluctuation effects, contrary to 
previous attempts, based on contour integration in the complex plane, that
required the limiting behavior of hydrodynamics at infinite frequencies 
\cite{Chan}. In fact, our calculation explicitly includes the scale at 
which the macroscopic 
hydrodynamics breaks down. The limit of vanishing compressibility is 
non-trivial and has to be taken carefully, because it can never be 
derived from a realistic inter-particle potential with infinite stiffness 
\cite{vankampen}.  

We interpret the non-zero hydrodynamic force correlations predicted in this work
as a modification of  the thermal stochastic force correlations that act on a 
Brownian particle in a fluid. Since the force correlator depends on the 
separation between the particles, the bath-mediated force fluctuations between 
the particles would modify the 
particles' Langevin dynamics and thus, in principle, should be detectable 
\cite{deBacco,Netz}. The separation dependence 
of 
the normal force cross-correlation function represents an interesting case of 
colloidal bodies which do not interact directly, but are driven by correlated 
noise sources that can provide an alternative mechanism which can produce 
non-trivial, ordered steady states \cite{Maritan}.  We have considered infinite 
bounding surfaces, so our results are not strictly applicable to the case of 
finite particles, but our calculation can be straightforwardly generalized to 
include a spherical geometry, which would also admit an analytic, albeit 
much more complicated, solution. Moreover, we intend to include the effects of 
heat transfer in a future calculation.

For experimental verification of our results, we again note that one would
have to generalize our calculation to the case of two spheres in a fluctuating 
hydrodynamic medium. This is different from the existing analysis of 
fluctuations of two unconnected, but hydrodynamically interacting spheres 
\cite{Netz}, a problem in some sense dual to ours. In order to exploit this 
connection, our first step will be to calculate the 
cross-correlation function for two spherical particles. 

\begin{acknowledgments}
This work has been partially funded by the U.S.~Department of Energy. 
C.M.~was supported in part by the 
U.S.~National Science Foundation under Grant NSF PHY10-034278. 
A.N.~acknowledges partial support from the 
Royal Society, the Royal Academy of Engineering, and the British Academy. 
B.-S.L. and R.P.~also acknowledge the financial support of the Agency for 
research and development of Slovenia (ARRS) under the bilateral SLO-A Grant 
No.~N1-0019. We acknowledge illuminating discussions with M.~Kardar in the KITP 
program on 
{\em The Theory and Practice of Fluctuation-Induced Interactions} (2008). 
R.P.~would like to thank 
Joel Cohen for his careful reading of the manuscript and for his comments.
\end{acknowledgments}

\appendix

\section{\label{app:f3f3simple}Derivation of the force correlator, 
Eq.~\eqref{eq:czzpt}}

In this Appendix, we derive the explicit expression, Eq.~\eqref{eq:czzpt}, for  
the force correlator defined in terms of the stress tensor in 
Eq.~\eqref{eq:forcecorr}. Our starting point is the general expression 
\begin{equation}
\big\langle {\cal F}_i^{(1)} (t){\cal F}_j^{(1)}(t) \big\rangle =
\iint_{\Gamma}\left\langle
\sigma_{ik}^{(1)}(\mathbf{r};t)\sigma_{jl}^{(1)}(\mathbf{r}';
t)\right\rangle\, \mathrm{d} A_k \mathrm{d} A_l',
\end{equation}
where repeated subindices are summed over.
In principle, this equation represents nine components of the force variance, 
each of which has nine contributions. For this work, we 
are interested in only the $i = j= z$ component of the force acting on the 
plane parallel to the boundaries. We thus have
\begin{align}\label{eq:fifjAPP}
{\cal C}(z,z';{} & t,t') = \big\langle {\cal F}_z^{(1)} (z;t){\cal 
F}_z^{(1)}(z';t') 
\big\rangle \nonumber\\
 = {} & 
 \iint_A\left\langle
\sigma_{zz}^{(1)}(\mathbf{r};t)\sigma_{zz}^{(1)}(\mathbf{r}^{\prime};
t')\right\rangle\, \mathrm{d} A_z \mathrm{d} A_z^{\prime},
\end{align}
where $A$ is the surface area for each of the plates and $\mathrm{d} A_z 
=\mathrm{d}x\mathrm{d}y$ and $\mathrm{d} 
A_z^{\prime}=\mathrm{d}x'\mathrm{d}y'$.
The first-order stress tensor is given by
\begin{align}\label{eq:sigma1APP}
\sigma_{ij}^{(1)}(\mathbf{r};t) = {} & \eta \left(\nabla_i
v_j^{(1)}(\mathbf{r};t) + \nabla_j v_i^{(1)}(\mathbf{r};t)\right) \\
{} &\hspace{-1.5cm} - 
\left[\left(\frac{2\eta}{3} - \zeta\right) \nabla_k
v_k^{(1)}(\mathbf{r};t) + 
c_0^2\rho^{(1)}(\mathbf{r};t)\right]\delta_{ij}+S_{ij}(\mathbf{r};t).\nonumber
\end{align}
In calculating the force correlator, which follows by inserting 
\eqref{eq:sigma1APP} into \eqref{eq:fifjAPP}, we realize that we are ultimately 
interested in these correlation functions evaluated at the boundaries with 
no-slip boundary conditions. Therefore, the terms that contain a 
derivative with respect to the transverse directions acting on the velocity 
field will vanish. On the other hand, the spatial (surface) integral over the 
transverse correlation function 
$c_{zz}^{\mathrm{T}\mathrm{T}}(\mathbf{r},\mathbf{r}';t,t')=\big\langle  
v_z^{\mathrm{T}}(\mathbf{r};t)v_z^{\mathrm{T}}(\mathbf{r}^\prime;t') 
\big\rangle$ also vanishes (see Appendix \ref{app:vtzero}). 
It is also straightforward to show that the terms containing cross 
correlations between the random stress tensor and other fluctuating fields 
vanish; this is because these terms turn out to be proportional to 
$\nabla_z\delta(z-z')$, which is zero for $z\neq z'$ and can also be set to 
zero for $z=z'$ by using a standard regularization scheme (e.g., by considering 
the 
Dirac delta function as a limiting form of a Gaussian function). Hence, the 
expression for the force correlator is:
\begin{align}
{\cal C}(z,z';{} &t,t') 
=  \iint_A\mathrm{d} A_z \mathrm{d} 
A_z'\bigg\{ \left\langle
S_{zz}(\mathbf{r};t)S_{zz}(\mathbf{r}^{\prime};t')\right\rangle \nonumber\\
{} & \;
+\left(\frac{4\eta}{3} +\zeta\right)^2
\nabla_z\nabla_z'\left\langle 
v_z^{(1)}(\mathbf{r};t)v_z^{(1)}(\mathbf{r}';t')\right 
\rangle\nonumber\\
{} & \; -\left(\frac{4\eta}{3} +\zeta\right) c_0^2\nabla_z\left\langle 
v_z^{(1)}(\mathbf{r};t)\rho^{(1)}(\mathbf{r}';t')\right 
\rangle\nonumber\\
{} & \; -\left(\frac{4\eta}{3} +\zeta\right) c_0^2\nabla_z'\left\langle 
\rho^{(1)}(\mathbf{r};t)v_z^{(1)}(\mathbf{r}';t')\right  
\rangle \nonumber\\
{} & \quad + c_0^4 \left\langle 
\rho^{(1)}(\mathbf{r};t)\rho^{(1)}(\mathbf{r}';t')\right 
\rangle\bigg\}.
\label{eq:czzpt_app_0}
\end{align}
Finally, the contributions from the correlation functions between 
the density and velocity fields (third and fourth terms in 
Eq.~\eqref{eq:czzpt_app_0})
 cancel out (see 
App.~\ref{app:pvzero}). Therefore, we find 
\begin{align}
{\cal C}(z,z';{} &t,t') =  {\cal P}_0(z,z';t,t') \nonumber\\
{} &+ \iint_A\mathrm{d} A_z \mathrm{d} 
A_z'\Big\{c_0^4 c^{\rho\rho}(\mathbf{r},\mathbf{r}';t,
t')  \nonumber\\
{} & \qquad\quad +\eta^2\chi^2
\nabla_z\nabla_z' 
c_{zz}^{\mathrm{LL}}(\mathbf{r},\mathbf{r}';t,
t')   
 \Big\}, 
 \label{eq:czzpt_app}
\end{align}
where ${\cal P}_0(z,z';t,t') =2k_{\mathrm{B}}T \eta \chi A 
\delta(z-z')\delta(t-t')$ and   
$\chi = (4/3+\zeta/\eta)$. This is nothing but Eq.~\eqref{eq:czzpt}.

\section{\label{app:vtzero}Transverse velocity correlator does not contribute}

The derivation of the correlation function for the transverse velocity fields 
largely follows that for the longitudinal components (Sec.~\ref{sec:corrfns}). 
In terms of the 
stochastic stress, the transverse velocity correlation function is
\begin{align}\label{eq:cijfreqAPP}
\big\langle  
v_i^{\mathrm{T}}(\mathbf{r};t) & {}
v_j^{\mathrm{T}}(\mathbf{r}^\prime;t') \big\rangle  = \int 
\mathrm{d}t^{\prime\prime} \int 
\mathrm{d}t^{\prime\prime\prime}\int 
\mathrm{d}^3\mathbf{r}^{\prime\prime} \,\int 
\mathrm{d}^3\mathbf{r}^{\prime\prime\prime} \nonumber\\
& {} \times
G^{\,\mathrm{T}}(\mathbf{r},\mathbf{r}^{\prime\prime};t-t^{\prime\prime})G^{\,
\mathrm { T
}}(\mathbf{r}^\prime,\mathbf{r}^{\prime\prime\prime};t'-t^{
\prime\prime\prime } )\nonumber\\
& {} \quad \times\left\langle
\Sigma^{\mathrm{T}}_i(\mathbf{r}^{\prime\prime};t^{\prime\prime})\Sigma^{\mathrm
{T } } _j(\mathbf {
r}^{\prime\prime\prime};t^{
\prime\prime\prime })\right\rangle.
\end{align}
Here, the transverse Green function satisfies
\begin{equation}
\left[\nabla_z^2 - q^2\right]\widetilde{G}^{\,\mathrm{T}}(z,z^{
\prime\prime}
;\mathbf{k};\omega)
= \frac{1}{\eta}\delta(z-z^{\prime\prime}),
\end{equation}
where $q^2 = (k^2-i\omega\rho_0/\eta)$. The solution for parallel-plane 
boundaries is
\begin{equation}
\widetilde{G}^{\,\mathrm{T}}(z,z^{
\prime\prime}
;\mathbf{k};\omega) = g_1^{\,\mathrm{T}}e^{-qz}+g_2^{\,\mathrm{T}}e^{q(z-L)} - 
\frac{1}{2 \eta q}e^{-q|z-z''|},
\end{equation}
with constants of integration given by
\begin{align}
g_1^{\,\mathrm{T}} = {} & \frac{1}{2q\eta} 
\csch(qL)\sinh(q(L-z'')), \\
g_2^{\,\mathrm{T}} = {}& \frac{1}{2q\eta}   \csch(qL)\sinh(qz'').
\end{align}

Recalling the stochastic properties of the stress tensor, which are
\begin{align}
\left\langle
\Sigma^{\mathrm{T}}_i(\mathbf{r};t)\,\Sigma^{
\mathrm{T}}_j(\mathbf{r}^{\prime};t') \right\rangle= {} & 
2k_{\mathrm{B}}T\eta\left(
\nabla_k\nabla_k^\prime \delta_{ij} -
\nabla_i\nabla_j^\prime
\right)\nonumber\\
 & \quad \times \delta (\mathbf{r}-\mathbf{r}^{\prime})\delta (t -
t')
\label{app:sigma_corr}
\end{align}
in the time domain, 
we can immediately carry out one of the time integrals and one of the 
the spatial integrals. Moreover, we are only 
concerned with the $i=j=z$ component, which leads to
\begin{align}
& c_{zz}^{\mathrm{T}\mathrm{T}}(\mathbf{r},\mathbf{r}';t,t')
=2 k_{\mathrm{B}}T\eta\int\mathrm{d}t''\int 
\mathrm{d}^3\mathbf{r}''
\\
& {} \quad \times \bigg\{
\nabla^{\prime\prime}_k  
G^{\,\mathrm{T}}(\mathbf{r},\mathbf{r}^{\prime\prime} 
;t-t'')\nabla^{\prime\prime}_k   
G^{\,\mathrm{T}}(\mathbf{r}^\prime,
\mathbf{r}^{\prime\prime};t'-t'') \nonumber\\
{} & \qquad - \nabla^{\prime\prime}_z  
G^{\,\mathrm{T}}(\mathbf{r},\mathbf{r}^{\prime\prime} 
;t-t'')\nabla^{\prime\prime}_z   
G^{\,\mathrm{T}}(\mathbf{r}^\prime,
\mathbf{r}^{\prime\prime};t'-t'')\bigg\}.\nonumber
\label{eq:vvtfrepAPP}
\end{align}
Integrating by parts, this becomes
\begin{align}
& c_{zz}^{\mathrm{T}\mathrm{T}} (\mathbf{r},\mathbf{r}';t,t')
=-2 k_{\mathrm{B}}T\eta\int\mathrm{d}t''\int 
\mathrm{d}^3\mathbf{r}''
\\
& {} \;\times 
\left[\left(\nabla^{\prime\prime\,2} - {\nabla^{\prime\prime}_z}^2\right)  
G^{\,\mathrm{T}}(\mathbf{r},\mathbf{r}^{\prime\prime} 
;t-t'')\right] 
G^{\,\mathrm{T}}(\mathbf{r}^\prime,
\mathbf{r}^{\prime\prime};t'-t'').\nonumber
\end{align}

Moving to the frequency representation and substituting the 
translation-invariant form of the Green function, analogous to
Eq.~\eqref{eq:Gdef}, we obtain
\begin{align}
 c_{zz}^{\mathrm{T}\mathrm{T}} {} &(\mathbf{r},\mathbf{r}';t,t') =
2k_{\mathrm{B}}T\eta
\int\frac{\mathrm{d}\omega'}{2\pi}e^{i 
\omega' 
(t-t')}\int 
\mathrm{d}z'' \int \frac{\mathrm{d}^2 
\mathbf{k}}{(2\pi)^2} \nonumber\\
& \times
e^{i\mathbf{k}\cdot
(\mathbf{s}-\mathbf{s}')}\mathbf{k}^2
\widetilde{G}^{\,\mathrm{T}}(z,z'' 
;\mathbf{k};-\omega') 
\widetilde{G}^{\,\mathrm{T}}(z',
z'';-\mathbf{k};\omega').
\end{align}
Here, we have integrated over $x''$ and $y''$, which generates a wavenumber 
delta function 
$\delta(\mathbf{k}+\mathbf{k}')$ that simplifies one of the 
wavenumber integrals.

We calculate the force variance, ${\cal C}(z,z';t,t')$, by integrating
the 
velocity correlation function over the boundaries of the fluid, 
i.e.~over $x$, $x'$, $y$, and $y'$ (see  Eq.~\eqref{eq:czzpt}). The only 
dependence on these variables occurs in the exponential function $
e^{i\mathbf{k}\cdot (\mathbf{s}-\mathbf{s}')}$. Thus, this integral generates a 
second wavenumber Dirac delta function, $\delta(\mathbf{k})$. It is now 
straightforward to see that the wavenumber integral vanishes: The product of 
the Green functions at $\mathbf{k} = \mathbf{0}$ is finite and consequently the 
factor of $\mathbf{k}^2$ ensures that the integral vanishes.

\section{\label{app:corrfn}Derivation of the density correlator}

Here, we calculate the correlation function of the density 
fields, $c^{\rho\rho}$, given in Eq.~\eqref{eq:ppcorr}. We start with the 
continuity 
equation, Eq.~\eqref{eq:lns2}, which can be written as 
\begin{equation}
\dot\rho^{(1)}(\mathbf{r};t) + \rho_0\nabla 
\cdot \mathbf{v}^{\mathrm{L}}(\mathbf{r};t) = 0,
\end{equation}
where the dot indicates a time derivative. From here we construct the 
correlation function
\begin{equation}
\label{eq:ppstart13}
\big\langle \dot\rho^{(1)}(\mathbf{r};t) 
\dot\rho^{(1)}(\mathbf{r}';t')  
\big\rangle = \rho_0^2~ \nabla_i {\nabla'}_j \left\langle 
v_i^{\mathrm{L}}(\mathbf{r};t) 
v_j^{\mathrm{L}}(\mathbf{r}';t') \right\rangle.
\end{equation}

By introducing Fourier components, we can cast the left-hand side of this 
equation into the form
\begin{align}
\label{eq:ppstart14}
\big\langle \dot\rho^{(1)}(\mathbf{r};t) 
\dot\rho^{(1)}(\mathbf{r}';t')  
\big\rangle = {} & \int \frac{\mathrm{d}\omega}{2\pi} \int 
\frac{\mathrm{d}\omega'}{2\pi}
 e^{-i (\omega t + \omega' t')} \nonumber\\
 {} & \times\big\langle 
\widetilde{\dot\rho}\!\!\phantom{\rho}^{(1)}(\mathbf{r};\omega)
\widetilde{\dot\rho}\!\!\phantom{\rho}^{(1)}(\mathbf{r}';\omega') 
\big\rangle.
\end{align}
The correlation function of the right-hand side of Eq.~\eqref{eq:ppstart13} 
is given by
\begin{align}\label{eq:cvvAPP}
c_{ij}^{\mathrm{L}\mathrm{L}} {} & (\mathbf{r},\mathbf{r}';t,t') 
= 2k_{\mathrm{B}}T\eta\chi \int\frac{\mathrm{d}\omega}{2\pi}  
\int\mathrm{d}\omega'e^{-i( \omega 
t+ \omega' 
t')}\\
& {} \times\delta(\omega+\omega')  \int 
\mathrm{d}^3\mathbf{r}''\nabla_i\widetilde{G}^{\,\mathrm{L}}(\mathbf{r},\mathbf{
r}
'';\omega)\nabla_j'\widetilde{G}^{\,
\mathrm { L
}}(\mathbf{r}',\mathbf{r}'';\omega' ). \nonumber
\end{align}

Combining Eqs.~\eqref{eq:ppstart13}, \eqref{eq:ppstart14} and
\eqref{eq:cvvAPP}, we write
\begin{align}
\big\langle {} & 
\widetilde{\dot\rho}\!\!\phantom{\rho}^{(1)}(\mathbf{r};\omega) 
\widetilde{\dot\rho}\!\!\phantom{\rho}^{(1)}(\mathbf{r}';\omega')  
\big\rangle =  4 \pi k_{\mathrm{B}}T \rho_0^2 \eta\chi \delta (\omega +
\omega') \nonumber\\
{} & \times \int 
\mathrm{d}^3\mathbf{r}''\nabla_i\nabla''_i  
\widetilde{G}^{\,\mathrm{L}}(\mathbf{r},\mathbf{r}'' 
; \omega)\nabla'_j\nabla''_j \widetilde{G}^{\,\mathrm{L}}(\mathbf{r}',
\mathbf{r}'';\omega') . 
\end{align}
Then, substituting this result into Eq.~\eqref{eq:ppstart14} gives
\begin{align}
\big\langle  {} & {\dot\rho}^{(1)}(\mathbf{r};t) 
{\dot\rho}^{(1)}(\mathbf{r}';t)  
\big\rangle = 4 \pi k_{\mathrm{B}}T \rho_0^2 \eta\chi \int 
\mathrm{d}\omega'e^{i\omega'(t-t')} \\
{} & \times  \int 
\mathrm{d}^3\mathbf{r}''\nabla_i \nabla^{
\prime\prime}_i  
\widetilde{G}^{\,\mathrm{L}}(\mathbf{r},\mathbf{r}'' 
; -\omega')\nabla'_j \nabla^{
\prime\prime}_j \widetilde{G}^{\,\mathrm{L}}(\mathbf{r}',
\mathbf{r}'';\omega').\nonumber
\end{align}
Several more steps are needed. First of all, we note that for the Fourier 
components of the density field,  we have
\begin{align}
\big\langle{} &  \widetilde{\rho}\!\!\phantom{\rho}^{(1)}(\mathbf{r};\omega) 
\widetilde{\rho}\!\!\phantom{\rho}^{(1)}(\mathbf{r}';\omega')  
\big\rangle  = - 4 \pi k_{\mathrm{B}}T \rho_0^2 \eta\chi\frac{\delta 
(\omega +
\omega')}{\omega \omega'}  \nonumber\\
{} & \times \int 
\mathrm{d}^3\mathbf{r}''\nabla_i \nabla^{
\prime\prime}_i  
\widetilde{G}^{\,\mathrm{L}}(\mathbf{r},\mathbf{r}'' 
; \omega)\nabla'_j \nabla^{
\prime\prime}_j \widetilde{G}^{\,\mathrm{L}}(\mathbf{r}',
\mathbf{r}'';\omega'),
\end{align}
and, therefore, we finally find
\begin{align}
{} & c^{\rho\rho}(\mathbf{r},\mathbf{r}';t,t') =  \frac{k_{\mathrm{B}}T}{\pi} 
\rho_0^2 \eta\chi
\int \frac{\mathrm{d}\omega'}{\omega'^2} e^{i\omega'(t-t')} \\
{} & \times \int 
\mathrm{d}^3\mathbf{r}''\nabla_i \nabla^{
\prime\prime}_i  
\widetilde{G}^{\,\mathrm{L}}(\mathbf{r},\mathbf{r}'' 
; -\omega')\nabla'_j \nabla^{
\prime\prime}_j \widetilde{G}^{\,\mathrm{L}}(\mathbf{r}',\nonumber
\mathbf{r}'';\omega').
\end{align}
Here, the relevant derivatives are given by
\begin{align}
\nabla_i \nabla^{
\prime\prime}_i   
\widetilde{G}^{\,\mathrm{L}}(\mathbf{r},\mathbf{r}'' {} & ; \omega') 
= 
\int \frac{\mathrm{d}^2 
\mathbf{k}}{(2\pi)^2}\,e^{i\mathbf{k}\cdot
(\mathbf{s}-\mathbf{s}'')} \nonumber \\
{} & \times \left(\mathbf{k}^2+\nabla_z 
\nabla''_z\right)  
\widetilde{G}^{\,\mathrm{L}}(z,z''; \mathbf{k} ;\omega').
\end{align}
The force variance, ${\cal C}(z,z';t,t')$, follows by integrating
the 
velocity correlation function over the boundaries of the fluid, 
i.e.~over $x$, $x'$, $y$, and $y'$ (see  Eq.~\eqref{eq:czzpt}). The only 
dependence on these variables occurs in the exponential function $
e^{i\mathbf{k}\cdot (\mathbf{s}-\mathbf{s}')}$ and consequently this integral 
generates a Dirac delta function over the 
transverse wavenumbers, $(2\pi)^2 A \delta(\mathbf{k}+\mathbf{k}')$.
This leads directly to Eq.~\eqref{eq:ppcorr}.

\section{\label{app:pvzero}Density-velocity cross-correlator does not contribute}

To calculate the density-velocity cross-correlator,  
$c_i^{\mathrm{L}\rho}(\mathbf{r},\mathbf{r}';t,t')$, we 
first 
construct the cross-correlation function
\begin{equation}
\label{eq:vpstart134}
\big\langle  v_i^{\mathrm{L}}(\mathbf{r};t)  
\dot\rho^{(1)}(\mathbf{r}',t')  
\big\rangle = - \rho_0 \nabla'_j  \left\langle   
v_i^{\mathrm{L}}(\mathbf{r};t) ~
 v_j^{\mathrm{L}}(\mathbf{r}';t') \right\rangle.
\end{equation}

Following a similar line of reasoning to that for the density-density 
correlation function, we can write 
\begin{align}
\label{eq:pvstart135}
\big\langle {} &  v_i^{\mathrm{L}}(\mathbf{r};\omega)  
\widetilde{\dot\rho}\!\!\phantom{\rho}^{(1)}(\mathbf{r}';\omega')  
\big\rangle = - 4 \pi k_{\mathrm{B}}T \rho_0 \eta\chi\delta (\omega +
\omega')
 \nonumber\\
{} &\times \int 
\mathrm{d}^3\mathbf{r}''\nabla^{
\prime\prime}_i  
\widetilde{G}^{\,\mathrm{L}}(\mathbf{r},\mathbf{r}'' 
; \omega)\nabla'_j \nabla^{
\prime\prime}_j \widetilde{G}^{\,\mathrm{L}}(\mathbf{r}',
\mathbf{r}'';\omega').
\end{align}
This equation leads to
\begin{align}
\big\langle {} &  v_i^{\mathrm{L}}(\mathbf{r};\omega)  
\widetilde{\rho}\!\!\phantom{\rho}^{(1)}(\mathbf{r}';\omega')  
\big\rangle =  4\pi i k_{\mathrm{B}}T \rho_0\eta\chi\frac{\delta 
(\omega 
+
\omega')}{\omega'}  \nonumber\\
{} &\times \int 
\mathrm{d}^3\mathbf{r}''\nabla''_i  
\widetilde{G}^{\,\mathrm{L}}(\mathbf{r},\mathbf{r}'' 
; \omega)\nabla'_j \nabla^{
\prime\prime}_j \widetilde{G}^{\,\mathrm{L}}(\mathbf{r}',
\mathbf{r}'';\omega'). \label{eq:vpstart136}
\end{align}
Thus, we obtain
\begin{align}
{} &c_i^{\mathrm{L}\rho} (\mathbf{r},\mathbf{r}';t,t')  =  i 
\frac{k_{\mathrm{B}}T}{\pi} 
\rho_0\eta\chi  \int 
\frac{{\mathrm{d}}\omega'}{\omega'}e^{i\omega'(t-t')}\nonumber\\
{} & \times \int 
\mathrm{d}^3\mathbf{r}''\nabla^{
\prime\prime}_i  
\widetilde{G}^{\,\mathrm{L}}(\mathbf{r},\mathbf{r}'' 
;- \omega')\nabla'_j  \nabla^{
\prime\prime}_j \widetilde{G}^{\,\mathrm{L}}(\mathbf{r}',
\mathbf{r}'';\omega').
\end{align}
We now introduce the translational invariant form of the Green function and 
take $i = z$. The double spatial integral generates a double wavenumber Dirac
delta function, giving
\begin{align}
c_z^{\mathrm{L}\rho} {} &(\mathbf{r},\mathbf{r}';t,t')  = i 
\frac{k_{\mathrm{B}}T}{\pi} 
\rho_0\eta\chi \int \frac{\mathrm{d}\omega'}{\omega'}e^{i\omega'(t-t')} 
\nonumber\\
{} & \times \int 
\mathrm{d}z''\int \frac{\mathrm{d}^2 
\mathbf{k}}{(2\pi)^2}\,e^{i\mathbf{k}\cdot
(\mathbf{s}-\mathbf{s}')}\nabla^{
\prime\prime}_z 
\widetilde{G}^{\,\mathrm{L}}(z,z'';\mathbf{k}
;- \omega') \nonumber\\
{} & \qquad \times
\left(\mathbf{k}^2+\nabla'_z \nabla''_z\right)  
\widetilde{G}^{\,\mathrm{L}}(z',z'';-\mathbf{k};\omega').
\end{align}

Now, following the line of reasoning of the previous section, the double 
integrals over $(x,y)$ and $(x^\prime,y^\prime)$ in the full contribution to 
the force correlator generate a Dirac delta function over the 
transverse wavenumbers, $(2\pi)^2 A \delta(\mathbf{k})$. Therefore, we can 
write this contribution in terms of 
the Green functions as
\begin{align}
& \iint_A\mathrm{d} A_z \mathrm{d} 
A_z'\,\nabla_zc_z^{\mathrm{L}\rho}(\mathbf{r},\mathbf{r}';t,
t') = -\frac{k_{\mathrm{B}}T}{\pi} \rho_0\eta\chi A\nonumber\\
& \times  \int
\frac{\mathrm{d}\omega'}{\omega'}\sin[\omega'(t-t')] \int 
\mathrm{d}z''  \nabla_z
 \nabla''_z \widetilde{G}^{\,\mathrm{L}}(z,z^{
\prime\prime}
;\mathbf{0};-\omega')\nonumber\\
{} &\qquad \times
\nabla_z' \nabla''_z 
\widetilde{G}^{\,\mathrm{L}}(z',z''
;\mathbf{0};\omega').\label{eq:p3zero}
\end{align}
Since this expression is odd with respect to changing $(z, t)$ to $(z', t')$ 
and 
vice versa, it follows that the contributions to the force correlator from the 
correlation functions between the density and velocity fields (third and fourth 
terms) in Eq.~\eqref{eq:czzpt_app_0} vanish, i.e., 
\begin{align}
\iint_A\mathrm{d} A_z \mathrm{d} 
A_z'\left[\nabla_zc_z^{\mathrm{L}\rho}(\mathbf{r},\mathbf{r}';t,t') 
+\nabla_z'c_z^{\mathrm{L}\rho}(\mathbf{r}',\mathbf{r};t',t)\right]=0.
\end{align}

\section{\label{app:wmderiv}Derivation of the frequency integrals}

In this Appendix, we derive Eqs.~\eqref{eq:final1} and \eqref{eq:final2} as 
well as Eqs.~\eqref{eq:wmt} and \eqref{eq:wmcrosst}.

\subsection{Same-plate force correlator, Eq.~\eqref{eq:final1}}

We start with the first contribution to 
the force correlation function, which is given in Eq.~\eqref{eq:p1zero}:
\begin{align}\label{eq:p1startAPP}
& {\cal P}_{1}(z, z'; t,t') =\frac{k_{\mathrm{B}}T}{\pi} \eta^3\chi^3 A \int 
\mathrm{d}\omega'\cos[\omega'(t-t')]
 \\
& \times \int 
\mathrm{d}z''
\nabla_z
\nabla^{
\prime\prime}_z \widetilde{G}^{\,\mathrm{L}}(z,z^{
\prime\prime}
;\mathbf{0};-\omega')\nabla_z' \nabla''_z \widetilde{G}^{\,\mathrm{L}}(z',z''
;\mathbf{0};\omega') .\nonumber
\end{align}
The explicit expressions for the derivatives appearing on the right-hand side of 
the above equation are
\begin{align}
 {} & \nabla_z\nabla_z''\widetilde{G}^{\,\mathrm{L}}(z,z^{
\prime\prime}
;\mathbf{0};-\omega) =\frac{-\omega }{2 [i 
\rho_0c_0^2-
\left(4\eta/3 + 
\zeta\right)\omega 
]} \\
{} & \; \times
\bigg[- 2\delta (z-z'')+\lambda^\ast 
e^{-\lambda^\ast |z-z''|}+\lambda^\ast \csch(\lambda^\ast L) 
\nonumber\\
{} & \;\;\times \left(\cosh(\lambda^\ast (L-z''))e^{
-\lambda^\ast z }
+\cosh(\lambda^\ast  
z'')e^{\lambda^\ast (z-L)}\right) \bigg], \label{eq:D1APP}\nonumber\\
{} & \nabla_z'\nabla_z''\widetilde{G}^{\,\mathrm{L}}(z',z^{
\prime\prime}
;\mathbf{0};\omega) = \frac{\omega }{2 [i 
\rho_0c_0^2+
\left(4\eta/3 + 
\zeta\right)\omega 
]}\\
{} & \quad\times \bigg[- 2\delta (z'-z'')+\lambda e^{-\lambda |z'-z''|}+\lambda 
\csch(\lambda L) \nonumber \\
{} & \qquad\times\left(\cosh(\lambda (L-z''))e^{-\lambda 
z'}+\cosh(\lambda  
z'')e^{\lambda (z'-L)}\right)\bigg],\label{eq:D2APP}\nonumber
\end{align}
where we have used the fact that $m = \lambda$ when $\mathbf{k}^2 = 0$ and we 
note that $\lambda^\ast = \lambda(-\omega)$.

Now, in principle, we could 
multiply together the results and integrate over $z''$. It is simpler, however, 
to look ahead a little. We know that, for the force variance at a single plate,
we will ultimately evaluate this correlation function at
$z = z'= 0$, so then these derivatives become
\begin{align}
 {} & \nabla_z\nabla_z''\widetilde{G}^{\,\mathrm{L}}(0,z^{
\prime\prime}
;\mathbf{0};-\omega) =\frac{-\omega }{2 [i 
\rho_0c_0^2-
\left(4\eta/3 + 
\zeta\right)\omega 
]} \nonumber\\
{} & \quad \times 
\bigg[-2\delta(0-z'') +\lambda^\ast e^{-\lambda^\ast z''}
+\lambda^\ast \csch(\lambda^\ast L) \nonumber \\
{} & \qquad \times \left(\cosh(\lambda^\ast (L-z''))
+\cosh(\lambda^\ast  
z'')e^{-\lambda^\ast L}\right)\bigg],\\
{} & \nabla_z'\nabla_z''\widetilde{G}^{\,\mathrm{L}}(0,z^{
\prime\prime}
;\mathbf{0};\omega) =  \frac{\omega }{2 [i 
\rho_0c_0^2+
\left(4\eta/3 + 
\zeta\right)\omega 
]} \nonumber\\
{} & \quad \times \bigg[ -2\delta(0-z'')+\lambda e^{-\lambda z''}+\lambda 
\csch(\lambda L)\nonumber \\
{} & \qquad \times \left(\cosh(\lambda (L-z''))+\cosh(\lambda  
z'')e^{-\lambda L}\right)\bigg]. \label{eq:g0zp}
\end{align}
Here, we have simplified the expressions using $|-z''| = z''$ for $z''$ in the 
range $[0,L]$. 

The key simplification now is to notice that we can collect together many of 
the exponential terms, which simplify to give
\begin{align}
 {} & \nabla_z\nabla_z''\widetilde{G}^{\,\mathrm{L}}(0,z^{
\prime\prime}
;\mathbf{0};-\omega) =\frac{-\omega }{ [i 
\rho_0c_0^2-
\left(4\eta/3 + 
\zeta\right)\omega 
]} \nonumber\\
{} & \quad \times \Big[\lambda^\ast 
\cosh[\lambda^\ast(L-z'')]\csch[\lambda^\ast L] 
-\delta(0-z'')\Big],\\
{} & \nabla_z'\nabla_z''\widetilde{G}^{\,\mathrm{L}}(0,z^{
\prime\prime}
;\mathbf{0};\omega) =  \frac{\omega }{ [i 
\rho_0c_0^2+
\left(4\eta/3 + 
\zeta\right)\omega 
]} \nonumber\\
{} & \quad \times \Big[\lambda \cosh[\lambda(L-z'')]\csch[\lambda L] 
-\delta(0-z'')\Big].\label{eq:gderivAPP}
\end{align}

Let us now use this result in our full expression, giving
\begin{align}\label{eq:p100}
{\cal P}_1{} & (0,0; t,t') = \frac{k_{\mathrm{B}}T}{\pi}\eta^3\chi^3  A \int 
 \frac{\mathrm{d}\omega\,\omega^2 \cos[\omega(t-t')]
}{\rho_0^2c_0^4+\left(4\eta/3 + \zeta\right)^2\omega^2} \nonumber\\
{} & \times 
\int_0^L  
\mathrm{d}z'' \Big[\lambda^\ast 
\cosh[\lambda^\ast(L-z'')]\csch[\lambda^\ast L] 
-\delta(0-z'')\Big] \nonumber\\
{} & \qquad \qquad \times \Big[\lambda \cosh[\lambda(L-z'')]\csch[\lambda L] 
-\delta(0-z'')\Big] \nonumber\\
= {} & \frac{k_{\mathrm{B}}T}{\pi}\eta^3\chi^3  A \int 
 \frac{\mathrm{d}\omega\,\omega^2 \cos[\omega(t-t')]
}{\rho_0^2c_0^4+\left(4\eta/3 + \zeta\right)^2\omega^2} \nonumber\\
{} & \; \times 
\bigg[\frac{|\lambda|^2\left(\lambda_{\mathrm{R}}\sin[2\lambda_{\mathrm{I}}L]
+\lambda_{\mathrm{I}}\sinh[2\lambda_{\mathrm{R}}L]
\right)
}{2\lambda_{\mathrm{I}}\lambda_{\mathrm{R}}\left(\cosh[2\lambda_{\mathrm{R}}L]
-\cos[2\lambda_{\mathrm{I}}L]
\right)} 
\nonumber\\
{} & \quad 
-\frac{2\left(\lambda_{\mathrm{I}}\sin[2\lambda_{\mathrm{I}}
L]
+\lambda_{\mathrm{R}}\sinh[2\lambda_{\mathrm{R}}L]
\right) }{\left(\cosh[2\lambda_{\mathrm{R}}L]
-\cos[2\lambda_{\mathrm{I}}L]
\right)}+\delta(0)\bigg].
\end{align}

We note that, in the $L\rightarrow \infty$ limit, this result reduces to the 
expression for the semi-infinite fluid, Eq.~\eqref{eq:p1slAPP}, an important 
cross-check of our results. We now express 
our result in terms of the dimensionless 
parameters 
$\ell_+$, $\ell_-$, $\chi$, $\gamma$, and $\tau = (t-t')/t_0$, the 
dimensionless 
variable 
$u$ and 
the 
function $f_m(u)$ (see Sec.~\ref{subsec:para}):
\begin{align}
{} &{\cal P}_1 (0,0; t,t') = \frac{k_{\mathrm{B}}T}{\pi} \rho_0 
c_0^2\chi^3\frac{ A }{L}
  2\int_0^{u_\infty} 
\mathrm{d}u \, f_0(u)\cos[u\tau] \nonumber\\
{} & \quad \times \left\{ \frac{1}{\cosh[2\ell_+]
-\cos[2\ell_-]} 
\bigg[\left(\frac{\ell^2}{2\ell_-} 
-2\ell_-\right) \sin[2\ell_-]
\right.
\nonumber\\
{} &\qquad  \left. 
+\left(\frac{\ell^2}{2\ell_+} 
-2\ell_+\right) \sinh[2\ell_+] \bigg]
 + L \delta(0)
\right\}.
\end{align}
Here, $\ell^2 = \ell_+^2+\ell_-^2$ and we have used the fact that the integrand 
is symmetric in the frequency to 
rewrite the region of integration over the positive real axis only, up to the
dimensionless cutoff, $u_\infty = \delta^2/a^2$. We now define
\begin{align}
 {} & {\cal W}_0(0,0;\tau) =   2\int_0^{u_\infty} 
\mathrm{d}u \, f_0(u)\cos[u\tau] \nonumber\\
{} & \quad \times \frac{1}{\cosh[2\ell_+]
-\cos[2\ell_-]} 
\bigg[  
\left(\frac{\ell^2}{2\ell_-} 
-2\ell_-\right) \sin[2\ell_-]
\nonumber\\
{} &\qquad  +\left(\frac{\ell^2}{2\ell_+} 
-2\ell_+\right) \sinh[2\ell_+] 
\bigg]
\end{align}
and
\begin{equation}
{\cal V}_0(0,0;\tau) = 2\int_0^{u_\infty} 
\mathrm{d}u \, f_0(u)\cos[u\tau],
\end{equation}
and thus we have
\begin{align}
{\cal P}_1 (0,0;{} &  t,t')=  \frac{k_{\mathrm{B}}T}{\pi} \rho_0 
c_0^2\chi^3\frac{ A }{L}  \nonumber\\
{} & \times 
\Big[{\cal W}_0(0,0;\tau)  +L {\cal V}_0(0,0;\tau) \delta(0)\Big].
\label{eq:p1APP}
\end{align}

Now, we turn to the second contribution, ${\cal P}_2(0,0; t,t')$. The 
derivatives 
with respect to $x$ and $y$ will ultimately 
bring down factors of $k_x$ and $k_y$. When we integrate over the spatial 
directions, the Dirac delta functions in wavenumber will remove these terms. The 
result is then directly related to the equation above, except that there is an 
extra denominator of $\omega^2$. We thus find
\begin{align}
{} &{\cal P}_2(0,0; t,t') =  \frac{k_{\mathrm{B}}T}{\pi}\rho_0 c_0^2 \chi 
\frac{ A }{L} 
2\int_0^{u_\infty} 
\mathrm{d}u \, f_2(u)\cos[u\tau] \nonumber\\
{} & \quad \times 
\left\{
\frac{1}{\cosh[2\ell_+]
-\cos[2\ell_-]} 
\bigg[\left(\frac{\ell^2}{2\ell_-} 
-2\ell_-\right) \sin[2\ell_-]
\right.
\nonumber\\
{} &\qquad  + \left.
\left(\frac{|\lambda|^2}{2\ell_+} 
-2\ell_+\right) \sinh[2\ell_+] \bigg] 
+ L \delta(0)
\right\},
\end{align}
or, in terms of the frequency integrals ${\cal W}_2$, Eq.~\eqref{eq:wmt}, and 
${\cal V}_2$, Eq.~\eqref{eq:vmt0}, we have
\begin{align}
{\cal P}_2(0,0;  t,t'){} & =  \frac{k_{\mathrm{B}}T}{\pi}\rho_0 c_0^2 \chi 
\frac{ A }{L} \nonumber\\
{} & \times \Big[{\cal W}_2(0,0;\tau)+L {\cal V}_2(0,0;\tau)\delta(0)\Big].
\label{eq:P_2_app1}
\end{align}
This, too, reduces to the 
expression for the semi-infinite fluid, Eq.~\eqref{eq:p2slAPP}, in the 
$L\rightarrow \infty$ limit.

Finally, putting together Eqs.~\eqref{eq:p1APP} and \eqref{eq:P_2_app1} and 
using Eq.~\eqref{eq:vmt}, we obtain Eq.~\eqref{eq:final1}.

\subsection{Cross-plate force correlator, Eq.~\eqref{eq:final2}}

Here we derive Eq.~\eqref{eq:final2}, our final integral expression for 
the cross-plate force correlator. We start from Eqs.~\eqref{eq:p1startAPP}, 
\eqref{eq:D1APP} and \eqref{eq:D2APP} again, but now we need to evaluate one of 
the derivatives at $z' = L$,
\begin{align}
{} & \nabla_z\nabla_z''\widetilde{G}^{\,\mathrm{L}}(0,z^{
\prime\prime}
;\mathbf{0};-\omega) = \frac{-\omega }{2 [i 
\rho_0c_0^2-
\left(4\eta/3 + 
\zeta\right)\omega 
]} \nonumber\\
{} & \quad\times 
\bigg[-2\delta(0-z'')+\lambda^\ast e^{-\lambda^\ast z''}
+\lambda^\ast \csch(\lambda^\ast L) \nonumber\\
{} & \qquad \times \left(\cosh(\lambda^\ast (L-z''))
+\cosh(\lambda^\ast  
z'')e^{-\lambda^\ast L}\right)\bigg],\\
{} & \nabla_z'\nabla_z''\widetilde{G}^{\,\mathrm{L}}(L,z^{
\prime\prime}
;\mathbf{0};\omega) = \frac{\omega }{2 [i 
\rho_0c_0^2+
\left(4\eta/3 + 
\zeta\right)\omega 
]} \nonumber\\
{} & \quad \times \bigg[-2\delta (L-z'')+\lambda e^{-\lambda (L-z'')} +\lambda 
\csch(\lambda L) \nonumber\\
{} & \qquad \times \left(\cosh(\lambda (L-z''))e^{-\lambda 
L}+\cosh(\lambda  
z'')\right) \bigg]. \label{eq:dzprL}
\end{align}
Once again we can simplify matters by writing
\begin{align}
{} & \nabla_z\nabla_z''\widetilde{G}^{\,\mathrm{L}}(0,z^{
\prime\prime}
;\mathbf{0};-\omega) =\frac{-\omega }{i 
\rho_0c_0^2-
\left(4\eta/3 + 
\zeta\right)\omega} \nonumber\\
{} & \;\times \Big[\lambda^\ast 
\cosh[\lambda^\ast(L-z'')]\csch[\lambda^\ast L] 
-\delta(0-z'')\Big],
\end{align}
but, in this case, Eq.~\eqref{eq:dzprL} becomes
\begin{align}
\nabla_z'\nabla_z''{} & \widetilde{G}^{\,\mathrm{L}}(L,z^{
\prime\prime}
;\mathbf{0};\omega) = \frac{\omega }{i 
\rho_0c_0^2+
\left(4\eta/3 + 
\zeta\right)\omega} \nonumber\\
{} & \quad \times \Big[\lambda 
\cosh[\lambda z'']\csch[\lambda L] 
-\delta(L-z'')\Big].
\end{align}
Therefore, the first contribution to the force correlator between the two 
plates 
is
\begin{align}\label{eq:pxcorrAPP}
 {} & {\cal P}_1(0,L; t,t')  =\frac{k_{\mathrm{B}}T}{\pi} \eta^3\chi^3  A  
\int 
\frac{\mathrm{d}\omega \,\omega^2 \cos[\omega(t-t')]
}{\rho_0^2c_0^4+\left(4\eta/3 + \zeta\right)^2\omega^2} \nonumber\\
{} & \times 
\int_0^L 
\mathrm{d}z'' \Big[\lambda^\ast 
\cosh[\lambda^\ast(L-z'')]\csch[\lambda^\ast L] 
-\delta(0-z'')\Big] \nonumber\\
{} & \qquad  \times\Big[\lambda 
\cosh[\lambda z'']\csch[\lambda L] 
-\delta(L-z'')\Big]\nonumber\\
{} & \qquad \qquad  =  \frac{k_{\mathrm{B}}T}{\pi} \eta^3\chi^3  A  
\int 
\frac{\mathrm{d}\omega \,\omega^2 \cos[\omega(t-t')]
}{\rho_0^2c_0^4+\left(4\eta/3 + \zeta\right)^2\omega^2} \nonumber\\
{} & \times 
\bigg[\frac{|\lambda|^2\left(\lambda_{\mathrm{R}}\cosh[\lambda_{\mathrm{R}}L]
\sin[\lambda_{\mathrm{I } } L ]
+\lambda_{\mathrm{I}}\cos[\lambda_{\mathrm{I } } L ]\sinh[\lambda_{\mathrm{R}}L]
\right)
}{\lambda_{\mathrm{I}}\lambda_{\mathrm{R}}\left(\cosh[2\lambda_{\mathrm{R}}L]
-\cos[2\lambda_{\mathrm{I}}L]
\right)} 
\nonumber\\
{} &\quad 
-\frac{4\left(\lambda_{\mathrm{I}}\cosh[\lambda_{\mathrm{R}}L]\sin[\lambda_{
\mathrm{I}}
L]
+\lambda_{\mathrm{R}}\cos[\lambda_{\mathrm{I}}L]\sinh[\lambda_{\mathrm{R}}L]
\right) }{\cosh[2\lambda_{\mathrm{R}}L]
-\cos[2\lambda_{\mathrm{I}}L]}\bigg].
\end{align}
We now express our result in terms of the dimensionless parameters as before, 
giving 
\begin{align}
{\cal P}_1{} & (0,L; t,t') = \frac{k_{\mathrm{B}}T}{\pi} 
\rho_0 c_0^2\chi^3 \frac{A  }{L}  2\int_0^{u_\infty} 
\mathrm{d}u\, f_0(u)\cos[u\tau] \nonumber\\
{} & \times \frac{1}{\cosh[2\ell_+]
-\cos[2\ell_-]} 
\bigg[\bigg(\frac{\ell^2}{\ell_-} - 4\ell_-\bigg)\cosh[\ell_+]
\sin[\ell_-] \nonumber\\
{} & \qquad + \bigg(\frac{\ell^2}{\ell_+} - 4\ell_+\bigg)\cos[\ell_-]
\sinh[\ell_+]\bigg].
\label{eq:P1_app_2}
\end{align}

As in the case 
of the same-plate force correlator, the other contribution, ${\cal 
P}_2(0,L; t,t')$, is very simply related to ${\cal P}_1(0,L; t,t')$. We can 
write  the result immediately as
\begin{align}
{\cal P}_2{} & (0,L; t,t') = \frac{k_{\mathrm{B}}T}{\pi} 
\rho_0 c_0^2 \chi\frac{A  }{L}  2\int_0^{u_\infty} 
\mathrm{d}u\, f_2(u)\cos[u\tau] \nonumber\\
{} & \times \frac{1}{\cosh[2\ell_+]
-\cos[2\ell_-]} 
\bigg[\bigg(\frac{\ell^2}{\ell_-} - 4\ell_-\bigg)\cosh[\ell_+]
\sin[\ell_-] \nonumber\\
{} & \qquad + \bigg(\frac{\ell^2}{\ell_+} - 4\ell_+\bigg)\cos[\ell_-]
\sinh[\ell_+]\bigg].
\label{eq:P2_app_2}
\end{align}
Now, putting together Eqs.~\eqref{eq:P1_app_2} and \eqref{eq:P2_app_2} and 
defining the frequency integral
\begin{align}
{} &{\cal W}_m (0,L;\tau)   = 2\int_0^{u_\infty} 
\mathrm{d}u\,f_m(u)\cos[u\tau] \nonumber\\
{} & \quad\times \frac{1}{\cosh[2\ell_+]
-\cos[2\ell_-]} 
\bigg[\bigg(\frac{\ell^2}{\ell_-} - 4\ell_-\bigg)\cosh[\ell_+]
\sin[\ell_-] \nonumber\\
{} & \qquad\qquad  + \bigg(\frac{\ell^2}{\ell_+} - 4\ell_+\bigg)\cos[\ell_-]
\sinh[\ell_+]\bigg],
\label{eq:wmcrosst_app} 
\end{align}
we obtain Eq.~\eqref{eq:final2}.

In the large plate-separation limit, $L\rightarrow \infty$, these 
correlators vanish, in accordance with the results of Appendix \ref{app:sinf}.

\section{\label{app:othergeom}Time correlators for simple geometries}

In this Appendix, we derive the time-dependent correlators for two simple geometries: A 
semi-infinite fluid with a single hard-wall boundary and an infinite fluid. The 
semi-infinite fluid  is the limiting case for the two-wall geometry
in the limit of infinite plate separation and we have confirmed, both analytically
and numerically, that our results for the two-wall geometry reduce to
the semi-infinite fluid results.

\subsection{\label{app:sinf}Semi-infinite fluid}

The Green function solution of Eq.~\eqref{eq:gl} for a semi-infinite fluid, 
with an infinite hard-wall boundary at $z=0$, is
\begin{equation}
\widetilde{G}^{\mathrm{L}}(z,z^{
\prime\prime}
;\mathbf{k};\omega) = \frac{i\lambda^2 }{2m 
\omega\rho_0}\left[e^{-m(z+z'')} - 
e^{-m|z-z''|}\right],
\end{equation}
where now $z'' > 0$. Once again, we substitute this result 
into Eqs.~\eqref{eq:p1zero} and \eqref{eq:p2zero}. The derivative we require 
this time is
\begin{align}\label{eq:semideriv}
\nabla_z\nabla_z''\widetilde{G}^{\mathrm{L}}(z,z'';\mathbf{k};\omega) = {} & 
\frac{i\lambda^2 }{2\omega\rho_0} \Big[me^{-m(z+z'')}+ me^{-m|z-z''|}
\nonumber\\
{} &  \qquad-2\delta(z-z'')\Big].
\end{align}

Carrying out the spatial integrals over $z''$, from zero to infinity, we obtain
\begin{align}
{\cal P}_{1}(0, 0; t,t') = {} & \frac{k_{\mathrm{B}}T}{\pi} \frac{\eta^3\chi^3 
A }{\rho_0^2}
\int \frac{\mathrm{d}\omega'}{\omega^{\prime\,2}} |\lambda|^4 
\cos[\omega'(t-t')]\nonumber\\
{} & \qquad \times\Big[\frac{|\lambda|^2}{2\lambda_{\mathrm{R}}} - 
2\lambda_{\mathrm{R}} + 
\delta(0)\Big],\label{eq:p1slAPP}\\
{\cal P}_2(0, 0; t,t') = {} & \frac{k_{\mathrm{B}}T}{\pi} \eta\chi 
c_0^4 
A \int
\frac{\mathrm{d}\omega'}{\omega^{\prime\,4}}|\lambda|^4 \cos[\omega'(t-t')] 
\nonumber\\
{} & \qquad \times\Big[\frac{|\lambda|^2}{2\lambda_{\mathrm{R}}} - 
2\lambda_{\mathrm{R}} + 
\delta(0)\Big] .\label{eq:p2slAPP}
\end{align}

We can carry out these frequency integrals for the equal time case, with $t=t'$, by first transforming to the 
dimensionless variables, $x = \eta\chi\omega/(\rho_0c_0^2)$ and $\tau_x = (t-t')\rho_0c_0^2/(\eta\chi)$, and then defining
$\tan \theta = x$. We obtain
\begin{align}
{} & {\cal P}_{1}(0, 0) = \frac{k_{\mathrm{B}}T}{\pi} 
2\rho_0c_0^2A\bigg\{\delta(0)\int_0^{x_\infty}
\frac{\mathrm{d}x\, x^2}{x^2+1}\nonumber\\
{} & \qquad\qquad \qquad +\frac{\rho_0c_0}{\sqrt{2}\eta\chi}\int_0^{\theta_\infty}
\mathrm{d}\theta\, \tan^3\theta\frac{\big(2\cos\theta - 1\big)}{\sqrt{\sec\theta - 1}}\bigg\}
\nonumber\\
{} &= \frac{k_{\mathrm{B}}T}{\pi} 
2\rho_0c_0^2A\bigg\{\delta(0)\big(x_\infty - \arctan x_\infty\big)\nonumber\\
{} & \quad +\frac{8\sqrt{2}\rho_0c_0}{3\eta\chi}
\frac{z_\infty(3-z_\infty)}{\sqrt{z_\infty-1}}\sin^4\left(\frac{1}{2}\arctan(z_\infty^2-1)\right)\bigg\},
\end{align}
and
\begin{align}
 {} &{\cal P}_{2}(0, 0) =  \frac{k_{\mathrm{B}}T}{\pi} 
2\rho_0c_0^2A\bigg\{\delta(0)\int_0^{x_\infty}
\frac{\mathrm{d}x}{x^2+1}\nonumber\\
{} & \quad +\frac{\rho_0c_0}{\sqrt{2}\eta\chi}\int_0^{\theta_\infty}
\mathrm{d}\theta\, \tan\theta\frac{\big(2\cos\theta - 1\big)}{\sqrt{\sec\theta - 1}}\bigg\}
\nonumber\\
{} &= \frac{k_{\mathrm{B}}T}{\pi} 
2\rho_0c_0^2A\bigg\{\delta(0)\arctan x_\infty+\frac{2\rho_0c_0}{\eta\chi}
\frac{\sqrt{z_\infty-1}}{z_\infty}\bigg\}.
\end{align}
Here, $\theta_\infty=\arctan x_\infty$ and $z_\infty=\sqrt{1+x_\infty^2}$ are both functions
of the dimensionless cutoff $x_\infty = \eta^2\chi/(a^2\rho_0^2c_0^2)$.

The equal-time correlator for a semi-infinite fluid is then given by
\begin{align}
{} & {\cal C}(0, 0) = \frac{k_{\mathrm{B}}T}{\pi} 
2\rho_0c_0^2A\bigg\{x_\infty \cdot\delta(0)+\frac{\rho_0c_0}{\eta\chi}\bigg[\frac{2\sqrt{z_\infty-1}}{z_\infty}\nonumber\\
{} & \; +\frac{8\sqrt{2}}{3}
\frac{z_\infty(3-z_\infty)}{\sqrt{z_\infty-1}}\sin^4\left(\frac{1}{2}\arctan(z_\infty^2-1)\right)\bigg]
\bigg\},
\end{align}
which is Eq.~\eqref{eq:c00si}.

The corresponding time-dependent, cross-plate force correlator vanishes for the 
semi-infinite fluid geometry, dropping to zero as $1/L$.

\subsection{Infinite fluid}

The Green function solution of Eq.~\eqref{eq:gl} for an infinite fluid, 
i.e., vanishing Green function at $z \rightarrow \pm \infty$, is
\begin{equation}
\widetilde{G}^{\mathrm{L}}(z,z^{
\prime\prime}
;\mathbf{k};\omega) = - 
\frac{i\lambda^2 }{2m 
\omega\rho_0}e^{-m|z-z''|},
\end{equation}
where $m^2 = \mathbf{k}^2+\lambda^2$ and $\lambda$ is the longitudinal decay 
constant defined in Eq.~\eqref{eq:lambda}. We will substitute this result 
into Eqs.~\eqref{eq:p1zero} and \eqref{eq:p2zero}. The derivative we require is
\begin{align}
\nabla_z\nabla_z''\widetilde{G}^{\mathrm{L}}(z,z'';\mathbf{k};\omega) = {} & 
\frac{i\lambda^2 }{2\omega\rho_0} \Big[me^{-m|z-z''|}
\nonumber\\
{} & \qquad \qquad-2\delta(z-z'')\Big].
\end{align}

For an infinite fluid, the spatial 
integral runs from negative infinity to positive infinity and we assume that we 
determine the time correlator at $z = z'=0$. Carrying out the spatial integral, 
we obtain
\begin{align}
{\cal P}_{1}(0, 0; t,t') = {} & \frac{k_{\mathrm{B}}T}{\pi} \frac{\eta^3\chi^3 
A }{\rho_0^2}
\int 
\frac{\mathrm{d}\omega'}{\omega^{\prime\,2}}\,|\lambda|^4\cos[\omega'(t-t')] 
\nonumber\\
{} & \qquad \times 
\bigg[\frac{|\lambda|^2}{4\lambda_{\mathrm{R}}}-\lambda_{\mathrm{R}}
+\delta(0)\bigg],\label{eq:p1zeroAPP}\\
{\cal P}_2(0, 0; t,t') = {} & \frac{k_{\mathrm{B}}T}{\pi} \eta\chi 
c_0^4 A \int
\frac{\mathrm{d}\omega'}{\omega^{\prime\,4}}|\lambda|^4 \cos[\omega'(t-t')] 
\nonumber\\
{} & \qquad \times 
\bigg[\frac{|\lambda|^2}{4\lambda_{\mathrm{R}}}-\lambda_{\mathrm{R}}
+\delta(0)\bigg].\label{eq:p2zeroAPP}
\end{align}

By comparing these results to Eqs.~\eqref{eq:p1slAPP} and \eqref{eq:p2slAPP},
we immediately see that this result is simply half that of the semi-infinite fluid.
Therefore, we have
\begin{equation}
{\cal C}(0, 0;t,t')\bigg|_{\mathrm{infinite}} = \frac{1}{2}{\cal C}(0, 0;t,t')\bigg|_{\mathrm{semi-infinite}}.
\end{equation}



\begin{thebibliography}{10}
\providecommand{\url}[1]{{#1}}
\providecommand{\urlprefix}{URL }
\expandafter\ifx\csname urlstyle\endcsname\relax
  \providecommand{\doi}[1]{DOI~\discretionary{}{}{}#1}\else
  \providecommand{\doi}{DOI~\discretionary{}{}{}\begingroup
  \urlstyle{rm}\Url}\fi


\bibitem{Casimir}
{H.B.G.~Casimir}, Proc.~K.~Ned.~Akad.~Wet.~\textbf{51}, 793 (1948).

\bibitem{Mostepanenko}
{V.M.~Mostepanenko} and {N.N.~Trunov}, \textit{The Casimir Effect and Its 
Applications} (Clarendon, Oxford, 1997).

\bibitem{Kardar}
{M.~Kardar} and {R.~Golestanian}, Rev.~Mod.~Phys.~\textbf{71}, 1233 (1999).

\bibitem{Bordag} {M.~Bordag}, {U.~Mohideen} and {V.M.~Mostepanenko}, 
Phys.~Rep.~\textbf{353}, 2 (2001).
  
\bibitem{Parsegian2005}
V.A. Parsegian, {\em Van der Waals Forces: A Handbook for Biologists, Chemists,  Engineers, and Physicists} (Cambridge University Press, 2005).

\bibitem{Mohideen} {M.~Bordag}, {G.L.~Klimchitskaya}, {U.~Mohideen} and 
{V.M.~Mostepanenko}, \textit{Advances in 
the Casimir Effect} (Oxford University Press, New York, 2009).
 
\bibitem{Dalvit} {D.A.R.~Dalvit}, {P.W.~Milonni}, {D.~Roberts}, {F.S.S.~Rosa}, {\em 
Casimir Physics}, Lecture Notes in Physics, Vol.~834, (Springer-Verlag, 
Berlin, 2011).

\bibitem{Krech} {M.~Krech}, \textit{The Casimir Effect in Critical Systems} 
(World Scientific, Singapore, 1994). 

\bibitem{French-RMP}
R. French et al., Rev. Mod. Phys. {\bf 82}, 1887 (2010).

\bibitem{Fisher} {M.E.~Fisher} and {P.G.~de Gennes}, C.~R.~Acad.~Sci.~Paris B 
\textbf{287}, 207 (1978).

\bibitem{Krech2}
{M.~Krech}, J.~Phys.~Condens.~Matter \textbf{11}, 391 (1999).

\bibitem{Krech3}
{M.~Krech}, Phys.~Rev.~E \textbf{56}, 1642 (1997).
  
\bibitem{Gambassi} 
{C.~Hertlein}, {L.~Helden}, {A.~Gambassi}, {S. Dietrich} and 
{C. Bechinger}, Nature \textbf{451}, 172 (2008).
  
\bibitem{Fukuto} 
{M.~Fukuto}, {Y.F.~Yano}, and {P.S.~Pershan}, Phys.~Rev.~Lett.~\textbf{94}, 
135702 (2005).

\bibitem{Schaeffer} {E.~Schaeffer} and {U.~Steiner}, Eur.~Phys.~J.~E 
\textbf{8}, 347 (2002).

\bibitem{Morariu04} {M.D.~Morariu}, {E.~Schaeffer} and {U.~Steiner},
Phys.~Rev.~Lett.~\textbf{92}, 156102 (2004).

\bibitem{Morariu03} {M.D.~Morariu}, {E.~Schaeffer} and {U.~Steiner},
Eur.~Phys.~J.~E \textbf{12}, 375 (2003).

\bibitem{LC}
{A.~Ajdari}, {L.~Peliti} and {J.~Prost}, Phys.~Rev.~Lett.~\textbf{66}, 1481 
(1991).

\bibitem{Li-kardar}
{H.~Li} and {M.~Kardar}, Phys.~Rev.~Lett.~\textbf{67}, 3275 (1991).

\bibitem{Antezza} {M.~Antezza}, {L.P.~Pitaevskii}, {S.~Stringari}, and 
{V.B.~Svetovoy}, Phys.~Rev.~A \textbf{77}, 022901 (2008).

\bibitem{Kruger} {M.~Kr\"uger}, {T.~Emig} and {M.~Kardar},
Phys.~Rev.~Lett.~\textbf{106} 201404 (2011).

\bibitem{Dean} {D.S.~Dean}, {V.A.~Parsegian}, and {R.~Podgornik}, 
Phys.~Rev.~A \textbf{87}, 032111 (2013).

\bibitem{Kirkpatrick13} {T.R.~Kirkpatrick}, {J.M.~Ortiz de Z\'arate}, 
{J.V.~Sengers}, Phys.~Rev.~Lett.~\textbf{110}, 235 (2013).

\bibitem{Kirkpatrick14} {T.R.~Kirkpatrick}, {J.M.~Ortiz 
de Z\'arate} and {J.V.~Sengers}, Phys.~Rev.~E \textbf{89}, 022145 (2014).

\bibitem{Ortiz} {J.M.~Ortiz de Z\'arate} and {J.V.~Sengers}, 
\textit{Hydrodynamic
Fluctuations in Fluids and Fluid Mixtures} (Elsevier, Amsterdam, 2006).

\bibitem{Lifshitz}
{E.M.~Lifshitz}, Sov.~Phys.~JETP \textbf{2}, 73 (1956).

\bibitem{rytov59}
{S.M.~Rytov}, \textit{Theory of Electric Fluctuations and Thermal Radiation}
(AFCRC-TR Air Force Cambridge Research Center, Bedford, 1959)

\bibitem{Rosa}
F.S.S. Rosa, D.A.R. Dalvit, and P.W. Milonni, 
Phys. Rev. A {\bf 81}, 033812 (2010).
 
\bibitem{Narayanaswami}
A. Narayanaswamy and Yi Zheng, 
Phys. Rev. A {\bf 88}, 012502 (2013).

\bibitem{Ajay1} D.S. Dean and A. Gopinathan, Phys.~Rev.~E {\bf 81}, 041126 (2010).

\bibitem{Ajay2} D.S. Dean and A. Gopinathan, J. Stat. Mech. L08001 (2009).

\bibitem{LandauLifshitz} {L.D.~Landau} and {E.M.~Lifshitz}, \textit{Statistical 
Physics, Part 2}, 1st Ed.~(Butterworth-Heinemann, Woburn, 1996).

\bibitem{Forster}
{D.~Forster}, \textit{Hydrodynamic Fluctuations, Broken Symmetry, and 
Correlation Functions} (W.A.~Benjamin, Reading, 1975). 

\bibitem{spohn}
H. Spohn, 
J. Phys. A: Math. Gen. {\bf 16}, 4275 (1983).

\bibitem{Tailleur}
J. Tailleur, J. Kurchan, and V. Lecomte, 
J. Phys. A: Math. Theor. {\bf 41}, 505001 (2008).
 
\bibitem{Rowlinson} {J.S.~Rowlinson}, {\em Cohesion:
A Scientific History of Intermolecular Forces} (Cambridge University Press, 
Cambridge, 2002).

\bibitem{Jones} {R.B.~Jones}, Physica A \textbf{105}, 395 (1981).
  
\bibitem{Chan} {D.Y.C.~Chan}, {L.R.~White}, Physica A \textbf{122}, 505 (1983).

\bibitem{vankampen}
N.G. van Kampen and J.J. Lodder, {\em Constraints}, Am. J. Phys. {\bf 52}, 419-424 (1984).
  
\bibitem{Dzyaloshinskii}
{I.E.~Dzyaloshinskii}, {E.M.~Lifshitz}, and {L.P.~Pitaevskii}, 
Sov.~Phys.~Uspekhi \textbf{4}, 153 (1961).
  
\bibitem{Ivlev} {B.I.~Ivlev}, J.~Phys.~Condens.~Matter \textbf{14}, 4829 (2002).
      
\bibitem{Bartolo} {D.~Bartolo}, {A.~Ajdari}, {J-B.~Fournier}, and 
{R.~Golestanian}, Phys.~Rev.~Lett.~\textbf{89}, 230601 (2002).

\bibitem{disorder-PRL}
A. Naji, D.S. Dean, J. Sarabadani, R. Horgan, R. Podgornik, Phys. Rev. Lett.
{\bf 104}, 060601 (2010).

\bibitem{pre2011}
D.S. Dean, A. Naji and R. Podgornik, Phys. Rev. E {\bf 83}, 011102 (2011).

\bibitem{epje2012}
A. Naji, J. Sarabadani, D.S. Dean and R. Podgornik, Eur. Phys. J. E {\bf 35},  24 (2012).

\bibitem{Note1}
We note the typographic error in Eq.~(2.1) of Ref.~\cite{Chan}, a missing occurrence of the field $\mathbf{v}$.
 
\bibitem{Note2}
Here we note a typographic error in Eq.~(2.11) of \cite{Chan}, corresponding to a factor of $1/\rho_0$ missing from Eq.~\eqref{eq:linLL2} above.

\bibitem{Erbas} {A.~Erbas}, {R.~Podgornik} and {R.R.~Netz}, Eur.~Phys.~J.~E 
\textbf{32}, 147 (2010).

\bibitem{Kim} {S.~Kim} and {S.~Karrila} \textit{Microhydrodynamics}, 1st 
Ed. (Dover, Mineola, 2005).
  
\bibitem{Schwinger} {J.~Schwinger}, {L.L.~Deraad Jr.}, {K.A.~Milton},
{W.~Tsai} and {J.~Norton}, \textit{Classical Electrodynamics} (Westview, 
Boulder, 1998). 

\bibitem{Raviv1} {U.~Raviv}, {P.~Laurat} and {J.~Klein}, Nature 
\textbf{413}, 5154 (2001).

\bibitem{Raviv2} {U.~Raviv} and {J.~Klein}, Science \textbf{297}, 1540 (2002).

\bibitem{Leng} Y.~Leng and P.T.~Cummings, Phys.~Rev.~Lett.~\textbf{94}, 026101 
(2005).

\bibitem{deBacco} {C.~De Bacco}, {F.~Baldovin}, {E.~Orlandini}, and 
{K.~Sekimoto}, Phys.~Rev.~Lett.~\textbf{112}, 180605 (2014).

\bibitem{Netz} J. von Hansen, A. Mehlich, B. Pelz, M. Rief, and R.R. Netz, Rev. 
Sci. Inst. {\bf 83}, 095116 (2012).

\bibitem{Maritan} S. Lise, A. Maritan and M.R. Swift, J. Phys. A: Math. Gen. {\bf 32}, 5251 (1999).


\end{thebibliography}
\end{document}